\documentclass[10pt,aps,showpacs,a4paper,nofootinbib]{revtex4}

\usepackage{graphicx}
\usepackage{graphics}
\usepackage{xspace}
\usepackage{amssymb}
\usepackage{amsmath}
\usepackage{latexsym}
\usepackage{mathrsfs}

\newcommand{\half}{\textstyle{\frac{1}{2}}}

\newcommand{\g}{\gamma}

\newcommand{\be}{\begin{equation}}
\newcommand{\ee}{\end{equation}}
\newcommand{\bea}{\begin{eqnarray}}
\newcommand{\eea}{\end{eqnarray}}
\newcommand{\nn}{\nonumber}
\newcommand{\nslash}{\kern 0.2 em n\kern -0.50em /}
\newcommand{\kslash}{\kern 0.2 em k\kern -0.45em /}
\newcommand{\pslash}{\kern 0.2 em p\kern -0.50em /}
\newcommand{\Sslash}{\kern 0.2 em S\kern -0.50em /}
\newcommand{\Pslash}{\kern 0.2 em P\kern -0.50em /}
\newcommand{\Rslash}{\kern 0.2 em R\kern -0.50em /}
\newcommand{\open}{{<\kern -0.3 em{\scriptscriptstyle )}}}
\newcommand{\rr}{|\vec{R}|}
\newcommand{\sst}{|\vec{S}_T|}
\newcommand{\lf}{\left}
\newcommand{\rg}{\right}
\newcommand{\h}{\hat{P}_{h\perp}}
\newcommand{\eps}{\epsilon}

\allowdisplaybreaks[2]

\begin{document}
\title{
Partial-wave analysis of two-hadron fragmentation functions
}

\author{Alessandro Bacchetta}
\email{alessandro.bacchetta@physik.uni-regensburg.de}
\altaffiliation{present address: 
Institut f\"ur Theoretische Physik, Universit\"at Regensburg,
D-93040 Regensburg, Germany}
\affiliation{Division of Physics and Astronomy, Faculty of Science, 
Free University, 
NL-1081 HV Amsterdam, the Netherlands}

\author{Marco Radici}
\email{radici@pv.infn.it}
\affiliation{Dipartimento di Fisica Nucleare e Teorica, 
Universit\`{a} di Pavia, 
and\\
Istituto Nazionale di Fisica Nucleare, Sezione di Pavia, I-27100 Pavia, Italy}

\begin{abstract}
We reconsider the option of extracting the transversity distribution by using 
interference fragmentation functions into two leading hadrons inside the same
current jet. To this end, we perform a new study of two-hadron fragmentation
functions. We derive new positivity bounds on them. 
We expand the hadron pair system in relative partial waves, so that we
can naturally incorporate in a unified formalism specific cases already studied in 
the literature, such as the fragmentation functions arising from the interference
between the $s$- and $p$-wave production of two mesons, as well as the production of 
a spin-one  hadron. In particular, our analysis clearly distinguishes two different ways 
to access the transversity distribution in two-hadron semi-inclusive leptoproduction.
\end{abstract}


\maketitle

\section{Introduction}
\label{sec:intro}

Two-hadron fragmentation functions have been proposed for the first time in 
Refs.~\cite{Collins:1994ax,Jaffe:1998hf} and then systematically analyzed at leading
twist in Ref.~\cite{Bianconi:1999cd}. 
The interest in these functions is mainly 
justified by the search for a mechanism to single out the chiral-odd transversity
distribution\footnote{See Ref.~\cite{Barone:2001sp} for a review on the topic.} 
in an alternative and technically simpler way than the Collins 
effect~\cite{Collins:1993kk}. In fact, in semi-inclusive deep 
inelastic scattering (SIDIS) where two unpolarized hadrons are produced in the 
current fragmentation region, i.e.\ for the reaction 
$ep \rightarrow e' h_1 h_2 {X}$, it is indeed possible to build a leading-twist 
single-spin asymmetry (SSA) containing the factorized product of the 
transversity 
and a chiral-odd two-hadron fragmentation function~\cite{Jaffe:1998hf,Radici:2001na}. 
In this process, the asymmetry occurs in the azimuthal angle between the
two-hadron plane and the laboratory plane;
the total momentum of the hadronic system does not need to have a
transverse component, i.e.\ out of collinearity with respect to 
the virtual photon axis. Therefore, the intrinsic transverse momentum 
of the quark can be integrated away and no transverse momentum dependent
functions are required, thus introducing 
simplifications both on the experimental and theoretical 
side~\cite{Boer:2001he} as compared to the Collins effect. 
Model calculations of such objects are feasible~\cite{Bianconi:1999uc} and
seem to produce measurable asymmetries in the SIDIS case~\cite{Radici:2001na}. 
Some of the two-hadron fragmentation functions are also naive time-reversal odd 
(T-odd) and originate from the interference between two production amplitudes with 
two different phases~\cite{Collins:1993kk,Bianconi:1999cd,Bacchetta:2001di}.
Therefore, in the literature these functions are usually referred to 
as interference fragmentation functions (IFF).

In an apparently independent context, semi-inclusive production of spin-1 
hadrons (e.g. $\rho$, $K^\ast$, $\phi$) has also been studied and proposed as 
a method to measure the transversity
distribution~\cite{Efremov:1982sh,Ji:1994vw,Anselmino:1996vq,Bacchetta:2000jk}.
To measure the polarization of the outgoing vector meson (e.g. $\rho^0$) it is 
necessary to measure the 4-momenta of the decay products (e.g. $\pi^+ \pi^-$). 
Thus, the reaction $ep \rightarrow e' \rho^0 {X} (\rho^0 \rightarrow \pi^+ \pi^-)$ is 
just a part of the more general reaction $ep \rightarrow e' \pi^+ \pi^- {X}$ 
(namely the part where the total invariant mass of the pion pair is equal to the 
$\rho$ mass). However, up to now the relation between spin-1 fragmentation functions
and two-hadron fragmentation functions has never been thoroughly examined, nor has ever 
been specified clearly how to access the transversity distribution in the case of spin-1 
fragmentation. The present work is motivated by the need to fill this gap.

Although in our work we focus mainly on SIDIS, two-hadron fragmentation
functions can be measured also in $e^+ e^-$ annihilation, if hadron pairs
belonging to the same jet are identified~\cite{Artru:1996zu,Marco}. Some data
are already available concerning two hadrons being produced via a spin-1 
resonance~\cite{Abbiendi:1999bz,Abreu:1995rg,Abreu:1997wd,Ackerstaff:1997kd,Ackerstaff:1997kj}.

The work is organized as follows. In Sec.~\ref{sec:2particle}, we will review the 
systematic analysis of semi-inclusive production of two unpolarized hadrons at leading 
twist. We will recover the results originally presented in 
Ref.~\cite{Bianconi:1999cd}. We will devote particular attention to the connection 
with the helicity basis formalism (see, e.g., 
Refs.~\cite{Jaffe:1996wp,Anselmino:1996vq}) and for the first time we will deduce 
positivity bounds on IFF.

In Sec.~\ref{sec:LM}, the whole problem is reconsidered by expanding in partial waves 
the two-hadron system in its center-of-mass frame. If we consider only low 
invariant masses, the expansion can be truncated to include the first two terms
only, as hadron pairs are produced mainly in the $s$-wave channel or in the $p$-wave 
channel (via a spin-1 resonance). We can thus deduce a general unifying formalism  
that naturally incorporates the specific case of Ref.~\cite{Jaffe:1998hf}, in the 
subsector describing the interference between relative $s$ and $p$ waves, as well as 
the case of spin-1 hadron fragmentation~\cite{Bacchetta:2000jk}, in the subsector of 
the relative $p$ wave. In particular, we will identify a SSA where the transversity 
distribution appears in connection with a $s$-$p$ IFF, and a SSA where the transversity 
is connected to a pure $p$-wave IFF. These two asymmetries are 
completely distinct, they could have different physical origins and different 
magnitudes.

In Sec.~\ref{sec:LM-T} we complete our analysis by including the intrinsic 
partonic transverse momentum and $\vec k_T$-unintegrated fragmentation functions. Also 
in this case, in Sec.~\ref{sec:LM-T2} we will present  positivity bounds and will carry 
out the partial wave expansion. The results for the complete cross section for all 
combinations of beam and target polarizations are listed in the appendices.
Finally, some conclusions are drawn in Sec.~\ref{sec:end}.


\begin{figure}[h]
\begin{center}
\includegraphics[height=5.cm, width=6.cm]{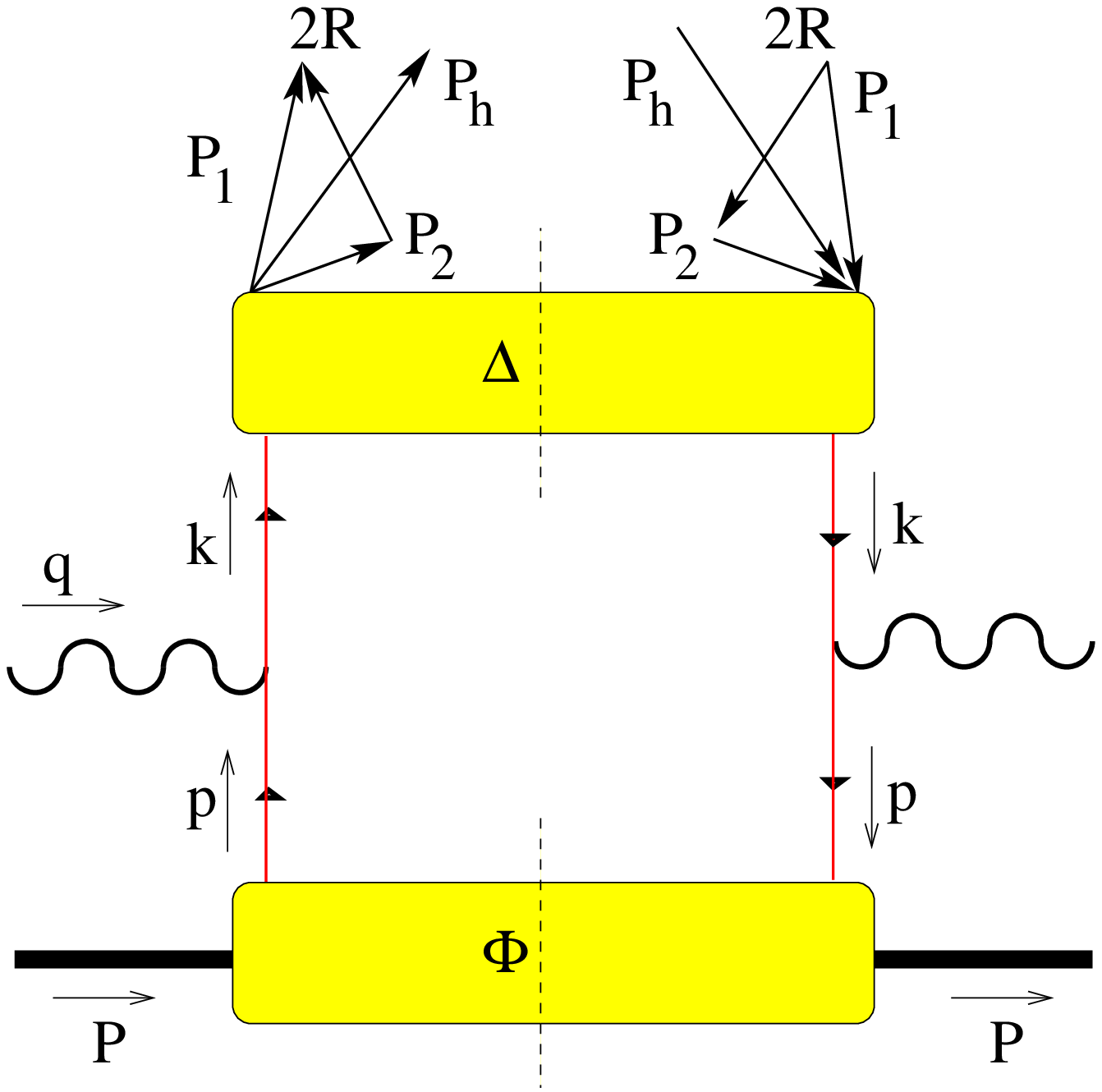}\hspace{2truecm}
\includegraphics[height=5.cm, width=6.cm]{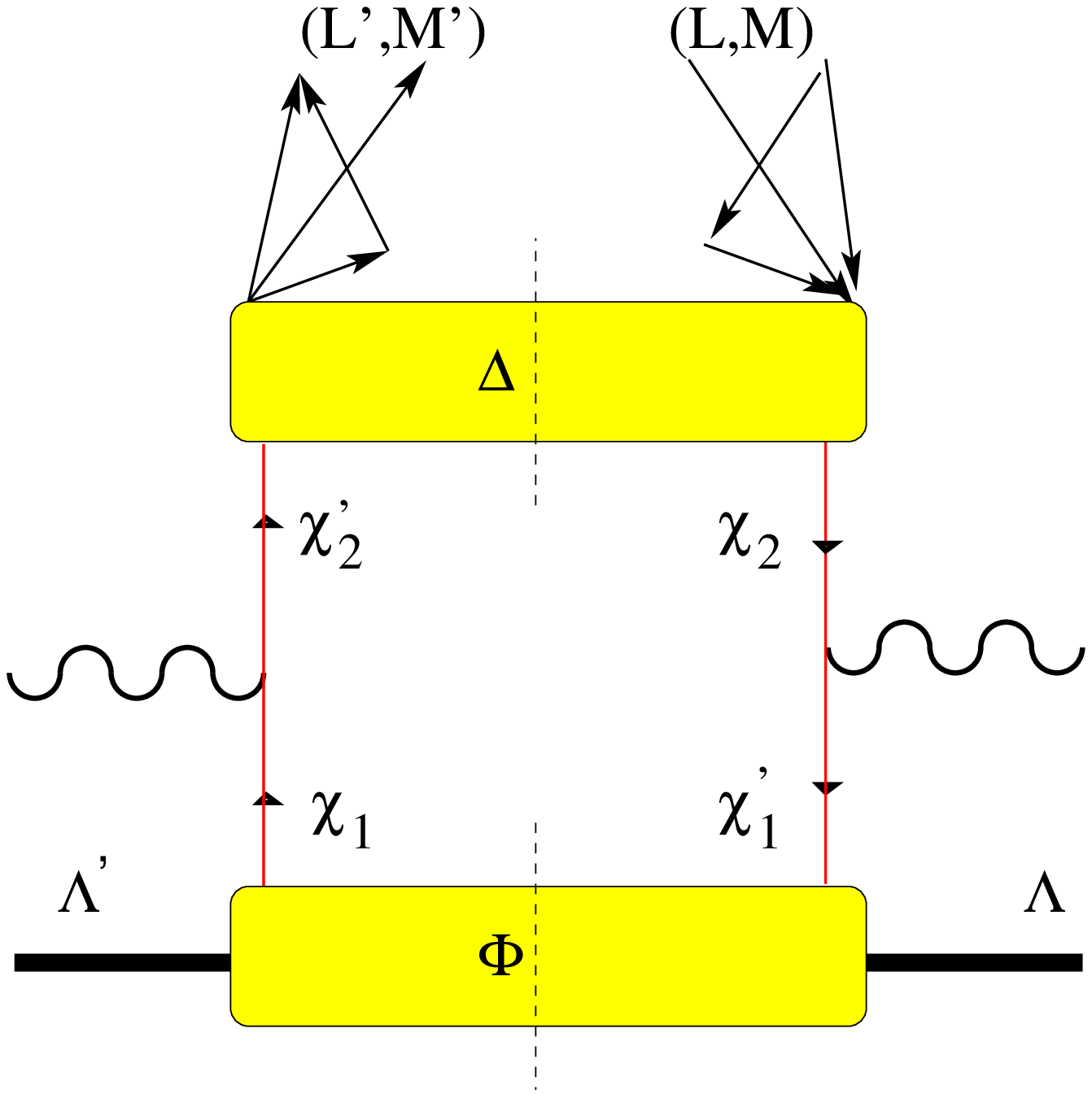}\\
a) \hspace{8truecm} b)
\end{center}
\caption {The usual quark handbag diagram contributing at leading twist to the
semi-inclusive DIS into two leading hadrons: a) hadron and parton momenta are shown, in
particular the total momentum $P_h=P_1+P_2$ and relative momentum $R=(P_1-P_2)/2$ of the
two-hadron system; b) target helicity, parton chirality and two-hadron partial wave
indices are shown.}
\label{fig:fig1}
\end{figure}


\section{Two-particle inclusive deep inelastic scattering}
\label{sec:2particle}

In the following, we will describe the kinematics and the details of the 
semi-inclusive production of two unpolarized hadrons in the context of the SIDIS process. 
However, we point out that the involved fragmentation functions can be 
used also in the case of reactions with a hadronic probe or in $e^+e^-$ 
annihilation~\cite{Artru:1996zu,Marco}. 


\subsection{Kinematics and hadronic tensor}
\label{sec:kin}

The process is schematically depicted in Fig.~\ref{fig:fig1}. An electron with 
momentum $l$ scatters off a target nucleon with mass $M$, polarization $S$ and 
momentum $P$, via the exchange of a virtual hard photon with momentum $q=l-l'$ 
($q^2 = -Q^2$). Inside the target, the photon hits a quark with momentum $p$, 
changing its momentum to $k=p+q$. The quark then fragments into a 
residual jet and two leading unpolarized hadrons with masses $M_1, M_2,$ and momenta 
$P_1$ and $P_2$. We introduce the vectors $P_h=P_1+P_2$ and $R=(P_1-P_2)/2$. We 
describe a 4-vector $a$ as $[a^-,a^+,\vec a_T ]$, i.e.\ in terms of its light-cone 
components $a^\pm = (a^0 \pm a^3)/\sqrt{2}$ and the bidimensional vector $\vec a_T$. 
It is convenient to choose the $\hat z$ axis according to the condition 
$\vec P_T = \vec P_{h\, T}=0$. In this case, the virtual photon has a nonvanishing
transverse momentum $\vec q_T$. However, it is also customary to align the $\hat z$ axis
opposite to the direction of the virtual photon, in which case the outgoing hadron has a
nonvanishing transverse momentum $\vec P_{h\perp} = -z\vec q_T$. These two directions
overlap up to corrections of order $1/Q$, which we will systematically neglect in the
following. The $y$ axis is chosen to point in the direction of 
the vector product $(-\vec{q} \times \vec{l}')$~\cite{Mulders:1996dh}.

We define the variables $x=p^+/P^+$, which represents the light-cone fraction of 
target momentum carried by the initial quark, and $z= P_h^-/k^-$, the light-cone 
fraction of fragmenting quark momentum carried by the final hadron pair. Analogously, 
we define the light-cone fraction $\zeta = 2R^-/P_h^-$, which describes how the total 
momentum of the hadron pair is split into the two single hadrons.\footnote{Note that 
$-1\leq \zeta \leq 1$, and $\zeta = 2\xi -1$, with $\xi$ defined in 
Ref.~\cite{Bianconi:1999cd}} The relevant momenta can be parametrized as
\bea
   P^\mu &= &\left[ \frac{M^2}{2P^+}, P^+, \vec 0 \right] \nn \\
   p^\mu &= &\left[ \frac{p^2+\vec p_T^{\,2}}{2xP^+}, xP^+, \vec p_T\right] \nn \\
   k^\mu &= &\left[ \frac{P_h^-}{z}, \frac{z(k^2+\vec k_T^2)}{2P_h^-},\vec k_T 
   \right] \nn \\
   P_h^\mu &= &\left[ P_h^-, \frac{M_h^2}{2 P_h^-}, \vec 0 \right] \nn \\
   R^\mu &= &\left[ \frac{\zeta}{2} P_h^-,
    \frac{(M_1^2-M_2^2)-\textstyle{\frac{\zeta}{2}}M_h^2}{2P_h^-}, \vec R_T
\right] \; .
\label{eq:kin}
\eea

Not all components of the 4-vectors are independent. In particular, here we observe 
that 
\bea
   R^2 &= &\frac{M_1^2+M_2^2}{2}\, - \, \frac{M_h^2}{4} \nn \\
   R_T^2 &= &\frac{1}{2}\, \left[ \frac{(1-\zeta)(1+\zeta)}{2} M_h^2 - (1-\zeta)
     M_1^2 - (1+\zeta) M_2^2 \right] \nn \\
   P_h \cdot R &= &\frac{M_1^2-M_2^2}{2} \nn \\
   P_h \cdot k &= &\frac{M_h^2}{2z} + z\, \frac{k^2+|\vec k_T|^2}{2} \nn \\
   R \cdot k &= &\frac{(M_1^2-M_2^2)-\textstyle{\frac{\zeta}{2}} M_h^2}{2z} + 
     z\zeta \, \frac{k^2+|\vec k_T|^2}{4} - \vec k_T \cdot \vec R_T \; .
\label{eq:kin2}
\eea
The positivity requirement $R_T^2 \ge 0$ imposes the further constraint
\begin{equation} 
    M_h^2 \ge \frac{2}{1+\zeta} M_1^2 + \frac{2}{1-\zeta} M_2^2  \; .
\label{eq:mh}
\end{equation} 

We shall first consider the case when the cross section is integrated over the 
transverse momentum of the virtual photon, $\vec{q}_T$, postponing the analysis 
of the complete case in Sec.~\ref{sec:LM-T}. Until then, no transverse-momentum dependent
distribution and fragmentation functions will appear. The seven-fold differential 
equation for two-particle-inclusive DIS is
\begin{equation}
   \frac{d^7\sigma}{d\zeta\;dM_h^2\;d\phi_R\;dz\;dx\;dy\;d\phi_S} = 
   \sum\limits_{a} \; \frac{\alpha^2 \,y\, e_a^2}{32\, z\, Q^4} \; L_{\mu\nu} \; 
   2M\,W_a^{\mu \nu} \; ,
\label{eq:cross}
\end{equation}
where $L_{\mu \nu}$ is the lepton tensor; $y=(E-E')/E$ is the fraction of beam energy 
transferred to the hadronic system and it is related to the lepton scattering 
angle in the center-of-mass (cm) frame; $\phi_R$ and $\phi_S$ are the azimuthal 
angles of $\vec R_T$ and $\vec S_T$ with respect to the lepton scattering plane. At tree 
level, the hadronic tensor for a flavour $a$ is given by 
\bea
   2M\, W_a^{\mu \nu} = 32 z \; \mbox{Tr} \big[ 
    \Phi_a(x,S) \, \gamma^\mu \, \Delta_a(z,\zeta,M_h^2,
    \phi_R) \, \gamma^\nu \big] + \left(\begin{array}{c} 
      q\leftrightarrow -q \\ \mu \leftrightarrow \nu
    \end{array} \right)\; ,
\label{eq:w}
\eea
where
\bea
   \Phi_a(x,S) &= &\int d \vec p_T\;d p^-\,\Phi_a(p;P,S) 
       \Big|_{p^+ = x P^+} \label{eq:phi} \\
   \Delta_a(z,\zeta,M_h^2,\phi_R) &= &\frac{z}{32}\int d \vec k_T \; 
       d k^+\,\Delta_a(k;P_h,R) \Big|_{k^- = P_h^-/z}  \; . \label{eq:delta}
\eea
The quark-quark correlator $\Phi$ describes the nonpertubative processes 
determining the distribution of parton $a$ inside the spin-1/2 target 
(represented by the lower shaded blob in Fig.~\ref{fig:fig1}) and, similarly, the 
correlator $\Delta$ symbolizes the fragmentation of quark $a$ producing two
tagged leading hadrons in a residual jet (upper shaded blob in Fig.~\ref{fig:fig1}). 


\begin{figure}[h]
\includegraphics[width=10.5cm]{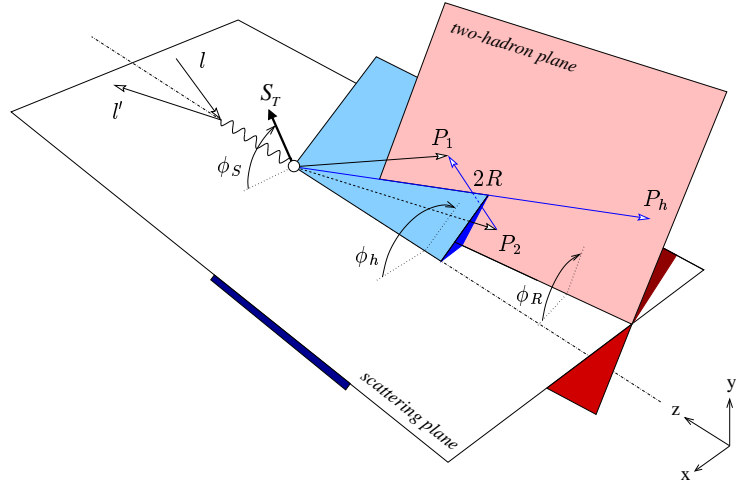} 
\caption {Kinematics for the SIDIS of the lepton $l$ on a transversely polarized 
target leading to two hadrons inside the same current jet.}
\label{fig:fig2}
\end{figure}


We are going to focus only on the leading twist contributions to the hadronic tensor 
of Eq.~(\ref{eq:w}). A method to extract these contributions consists in projecting 
the socalled good light-cone components out of the quark field $\psi$. 
As it is evident from the kinematics in the 
infinite momentum frame, the $+$ and the $-$ light-cone components are the dominant ones 
for the parton entering and exiting the hard vertex, respectively. They can be projected
out by means of the operators ${\mathcal P}_\pm = \half \gamma^\mp \gamma^\pm$. 
Any other component of $\psi$ is automatically of higher twist. Therefore, 
the hadronic tensor~(\ref{eq:w}) at leading twist looks like
\bea
  2M\,W_a^{\mu\nu} &= &32 z \, \mbox{Tr} \left[ {\mathcal P}_+ \Phi_a(x,S) \bar 
  {\mathcal P}_+ \, \gamma^\mu \, {\mathcal P}_- \Delta_a(z,\zeta,M_h^2,\phi_R) \bar 
  {\mathcal P}_- \, \gamma^\nu \right] \nn \\
  &= &32 z \, [{\mathcal P}_+ \Phi_a(x,S) \gamma^+]_{ij} \; [\half
  \gamma^- \gamma^\mu {\mathcal P}_-]_{jl} \; [\half \gamma^+ \gamma^\nu 
  {\mathcal P}_+]_{mi} \; [{\mathcal P}_- \Delta_a(z,\zeta,M_h^2,\phi_R)
  \gamma^-]_{lm} \; ,
\label{eq:wh}
\eea
where $\bar {\mathcal P}_\pm \equiv \gamma^0 {\mathcal P}_\pm^\dagger
\gamma^0$. In the last step the Dirac indices have been explicitly indicated. In the 
following, we will analyze each contribution to Eq.~(\ref{eq:wh}) separately.


\subsection{The quark-quark correlator $\Phi$}
\label{sec:phi}

The leading-twist projection of the quark-quark correlator $\Phi$ can be 
parametrized in terms of the well known distribution
functions~\cite{Ralston:1979ys,Bacchetta:1999kz}\footnote{Other common
notations are $f^a_1(x)=a(x)$, $g^a_1(x)=\Delta a(x)$, $h^a_1(x)=\delta a(x),
\Delta_T a(x)$~\cite{Barone:2001sp}.}
\bea
   {\mathcal P}_+ \Phi_a(x,S) \gamma^+ &= &
    \left( f^a_1(x) + \lambda g^a_1(x) \gamma_5 + 2h^a_1(x) \gamma_5 \Sslash_T 
    \right) \, {\mathcal P}_+ \nn \\[5pt]
   &= &\left( \begin{array}{cccc} 
              f^a_1 + \lambda g^a_1 & 0 & 0 & (S_x-iS_y) h^a_1 \\
	      0 & 0 & 0 & 0 \\
	      0 & 0 & 0 & 0 \\
	      (S_x+iS_y) h^a_1 & 0 & 0 & f^a_1 -\lambda g^a_1 \end{array} \right) \; ,
\label{eq:phihtw}
\eea
where $\lambda = M S^+/P^+$ and $\vec S_T=(S_x,S_y)$ are the light-cone helicity and
transverse components of the target spin, respectively 
(${\mathcal P}_+ \Phi$ corresponds to the $\vec p_T$-integrated 
parametrization of Eq.~(2) in Ref.~\cite{Radici:2001na}). It is possible to rewrite 
Eq.~(\ref{eq:phihtw}) in a more compact notation, namely in the chiral basis of the 
good quark fields $\psi_{\pm\,R/L} = {\mathcal P}_\pm\, {\mathcal P}_{R/L}\psi$, 
with ${\mathcal P}_{R/L} = (1\pm\gamma_5)/2$~\cite{Bacchetta:1999kz},
\begin{equation}
  [{\mathcal P}_+ \Phi_a(x,S) \gamma^+]_{\chi_1' \chi^{}_1} = \left( 
    \begin{array}{cc}
    f_1^a(x) + \lambda g_1^a(x) & (S_x-iS_y) h_1^a(x) \\[5pt]
    (S_x+iS_y) h_1^a(x) & f_1^a(x) - \lambda g_1^a(x) \end{array} \right) \; .
\label{eq:phictw}
\end{equation}
Finally, it is useful to project out also the target helicity density matrix
$\rho_{\Lambda \Lambda'}$ by
\begin{equation}
   [{\mathcal P}_+ \Phi_a \gamma^+]_{\chi_1' \chi^{}_1} = \rho_{\Lambda
   \Lambda'} \, [{\mathcal P}_+ \Phi_a
   \gamma^+]_{\chi_1' \chi^{}_1}^{\Lambda'\Lambda} \; ,
\label{eq:wctw2}
\end{equation}
with
\bea
   \rho_{\Lambda\Lambda'} &= &\frac{1}{2}\, \left( 
     \begin{array}{cc}
     1+\lambda & S_x-iS_y \\[2pt]
     S_x+iS_y & 1-\lambda \end{array} \right) ,\label{eq:rho} \\[5pt]
   [{\mathcal P}_+ \Phi_a \gamma^+]_{\chi_1'\chi^{}_1}^{\Lambda'\Lambda} 
      &= &\left( \begin{array}{cc|cc}
              f_1^a+g_1^a & 0 & 0 & 0 \\[2pt]
	      0 & f_1^a - g_1^a & 2h_1^a & 0 \\[2pt]
	      \hline 
	      0 & 2h_1^a & f_1^a - g_1^a & 0 \\[2pt]
	      0 & 0 & 0 & f_1^a + g_1^a \end{array} \right) \; . \label{eq:phihctw} 
\eea
In Eq.~(\ref{eq:phihctw}) the pair of indices $(\Lambda,\Lambda')$ identifies each
component of the $2\times2$ submatrices and indicates the spin state of the target; 
they are attached to each corresponding nucleon leg in the diagram of 
Fig.~\ref{fig:fig1}b. The pair $(\chi^{}_1,\chi_1')$ identifies each submatrix and 
indicates the parton chirality; they are attached to the emerging quark legs in 
Fig.~\ref{fig:fig1}b. Equation~(\ref{eq:phihctw}) satisfies general requirements, such as 
the angular momentum conservation $(\chi_1+\Lambda=\chi_1'+\Lambda')$, Hermiticity and 
parity invariance. The chiral transposed matrix is also positive semidefinite, from 
which the well known Soffer bound~\cite{Soffer:1995ww}, among others, is obtained:
\bea
   f_1^a(x) &\geq &0 \qquad f_1^a(x) \geq |g_1^a(x)| \nn \\
   |h_1^a(x)| &\leq &\half [f_1^a(x)+g_1^a(x)] \; .
\label{eq:dfbounds}
\eea


\subsection{The quark-quark correlator $\Delta$ and positivity bounds}
\label{sec:delta}

The most general parametrization of the quark-quark correlator $\Delta(k,P_h,R)$ entering 
Eq.~(\ref{eq:delta}), compatible with 
Hermiticity and parity invariance, is~\cite{Bianconi:1999cd} 
\bea
   \Delta(k,P_h,R) &= &M_h C_1 I + C_2 \Pslash_h + C_3 \Rslash + C_4 \kslash \nn \\
   & &+ \frac{C_5}{M_h} \sigma_{\mu \nu} P_h^\mu k^\nu + \frac{C_6}{M_h} \sigma_{\mu
   \nu} R^\mu k^\nu + \frac{C_7}{M_h} \sigma_{\mu \nu} P_h^\mu R^\nu \nn \\
   & &+ \frac{C_8}{M_h^2} \gamma_5 \varepsilon^{\mu\nu\rho\sigma} \gamma_\mu
   P_{h\,\nu} R_\rho k_\sigma \; ,
\label{eq:deltagen}
\eea
where the amplitudes $C_i(k^2,k\cdot P_h,k\cdot R,R^2)$ are dimensionless real scalar
functions. By using Eqs.~(\ref{eq:deltagen},\ref{eq:delta}) the
leading-twist projection becomes
\begin{eqnarray} 
   {\mathcal P}_- \Delta_a(z,\zeta,M_h^2,\phi_R) \gamma^- &= &
    \frac{1}{8 \pi} \, \left( D^a_1(z,\zeta,M_h^2) + i H_1^{\open\,a} (z,\zeta,M_h^2) 
    \frac{\Rslash_T}{M_h} \right) \, {\mathcal P}_- \nn \\[5pt]
   &= &\frac{1}{8 \pi} \, \left( \begin{array}{cccc} 
           0 & 0 & 0 & 0 \\
	   0 & D^a_1 & i e^{i\phi_R} \frac{|\vec R_T|}{M_h} H_1^{\open\,a} & 0 \\
	   0 & -i e^{-i\phi_R} \frac{|\vec R_T|}{M_h} H_1^{\open\,a} & D_1^a & 0 \\
	   0 & 0 & 0 & 0 \end{array} \right) \; ,
\label{eq:deltahtw}
\end{eqnarray} 
where 
\bea
   D_1(z,\zeta,M_h^2) &= &\frac{z \pi}{4}\, \int d^2\vec k_T dk^2 d(2k\cdot P_h) \,
   \delta \left( \vec k_T^2 + k^2 + \frac{M_h^2}{z^2} - \frac{2k\cdot P_h}{z}
   \right) \, \left[ C_2 + \frac{\zeta}{2} C_3 + \frac{1}{z} C_4 \right] 
\label{eq:d1} \\
   H_1^{\open}(z,\zeta,M_h^2) &= &\frac{z \pi}{4}\, \int d^2\vec k_T dk^2 
   d(2k\cdot P_h) \, \delta \left( \vec k_T^2 + k^2 + \frac{M_h^2}{z^2} - 
   \frac{2k\cdot P_h}{z} \right) \, \left[ \frac{1}{z} C_6 - C_7 \right]  \; .
\label{eq:h1}
\eea
The prefactors have been chosen to have a better connection with the
one-hadron results, i.e.\ after integrating over $\zeta$, $M_h^2$ and $\phi_R$.
In Eq.~(\ref{eq:deltahtw}), ${\mathcal P}_- \Delta$ corresponds to 
the parametrization of Eq.(3) in Ref.~\cite{Radici:2001na}.

The fragmentation function $H_1^{\open}$ is chiral odd and represents a possible partner 
to isolate the transversity distribution inside the cross section at leading 
twist~\cite{Radici:2001na}. Moreover, it is also odd with respect to naive 
time-reversal transformations (for brevity, T-odd)~\cite{Bianconi:1999cd}.
Noteworthy, it is the only example of leading-twist T-odd function surviving the 
integration upon the quark transverse momentum $\vec k_T$. It would be interesting to 
investigate it in order to understand what is the relevance of the transverse-momentum
dependence in generating T-odd effects~\cite{Brodsky:2002cx,Belitsky:2002sm}. 
As a consequence, the $\vec k_T$-integrated $H_1^{\open}$ could have simpler evolution
equations than the ones of the Collins function. Since $H_1^{\open}$ 
has the same operator structure as the transversity, it has been suggested that it 
could have the same evolution equations~\cite{Boer:2001zr,Stratmann:2001pt,Boer:2001zw}. 
However, the situation is complicated by the presence of two 
hadrons~\cite{Sukhatme:1980vs}\footnote{We thank D.~Boer for pointing out this detail.}, 
except for the component of $H_1^{\open}$ describing the production of a spin-1 resonance 
(see Sec.~\ref{sec:crosslm}).

Again, in the chiral basis for the good light-cone components Eq.~(\ref{eq:deltahtw}) 
is simplified to 
\begin{equation}
  [{\mathcal P}_- \Delta_a(z,\zeta,M_h^2,\phi_R) \gamma^-]_{\chi_2' 
  \chi_2^{}} = \frac{1}{8 \pi} \, \left( \begin{array}{cc}
       D_1^a(z,\zeta,M_h^2) & i e^{i\phi_R} \frac{|\vec R_T|}{M_h} 
       H_1^{\open\,a}(z,\zeta,M_h^2) \\[5pt]
       -i e^{-i\phi_R} \frac{|\vec R_T|}{M_h} H_1^{\open\,a}(z,\zeta,M_h^2) & 
       D_1^a(z,\zeta,M_h^2)  \end{array} \right) \; ,
\label{eq:deltactw}
\end{equation}
where $(\chi_2^{},\chi_2')$ are the quark chiralities to be attached to the parton
legs entering the $\Delta$ blob in Fig.~\ref{fig:fig1}b. 

The matrix in Eq.~(\ref{eq:deltactw}) is positive semi-definite, from which the 
following bounds can be derived:
\begin{equation}
   D_1^a(z,\zeta,M_h^2) \geq 0 \qquad D_1^a(z,\zeta,M_h^2) \geq 
   \frac{|\vec R_T|}{M_h} |H_1^{\open\,a}(z,\zeta,M_h^2)| \; .
\label{eq:ffbounds}
\end{equation}


\subsection{Cross section and transverse spin asymmetry}
\label{sec:cross}

Using the previous results, we can now rewrite the leading-twist cross section for
unpolarized two-hadron SIDIS in the helicity basis. In fact, after inserting 
Eqs.~(\ref{eq:wctw2}) and (\ref{eq:deltactw}) inside Eq.~(\ref{eq:wh}), the
cross section in
Eq.~(\ref{eq:cross}) becomes
\begin{equation}
 \frac{d^7\sigma}{d\zeta\;dM_h^2\;d\phi_R\;dz\;dx\;dy\;d\phi_S} = 
 \sum\limits_{a} \; \rho_{\Lambda\Lambda'}(S) \, 
 [{\mathcal P}_+ \Phi_a(x) \gamma^+]_{\chi_1'\chi_1^{}}^{\Lambda'\Lambda}
 \; \left( \frac{d\sigma^{eq_a}}{dy} \right)^{\chi_1^{}\chi_1'\,;\,\chi_2^{}\chi_2'} \; 
 [{\mathcal P}_- \Delta_a(z,\zeta,M_h^2,\phi_R) 
 \gamma^-]_{\chi_2' \chi_2^{}}  
\label{eq:crossh}
\end{equation}
where
\bea
   \left( \frac{d\sigma^{eq_a}}{dy} \right)^{\chi_1^{}\chi_1'\,;\,\chi_2^{}\chi_2'} 
   &= & \frac{e_a^2 \,\alpha^2 y}{Q^4} \, L_{\mu \nu} \, \left( 
   \frac{\gamma^- \gamma^\mu}{2} {\mathcal P}_- \right)^{\chi_1^{}\chi_2^{}} \; 
   \left( \frac{\gamma^+ \gamma^\nu}{2} {\mathcal P}_+ \right)^{\chi_2'\chi_1'} \nn
   \\
   &= &\frac{2 e_a^2 \, \alpha^2}{Q^2 y} \; \left( \begin{array}{cc|cc}
              A(y)+\lambda_e \frac{C(y)}{2} & 0 & 0 & -B(y) \\[2pt]
	      0 & 0 & 0 & 0 \\ \hline 
	      0 & 0 & 0 & 0 \\[2pt]
	      -B(y) & 0 & 0 & A(y)-\lambda_e \frac{C(y)}{2} \end{array} \right) 
\label{eq:eqcross}
\eea
represents the elementary electron-quark scattering. Strictly speaking, this is not a
scattering matrix, but a scattering amplitude times the conjugate of a 
different scattering amplitude~\cite{Anselmino:1996vq}. 
However, for conciseness we follow the 
notation of Ref.~\cite{Jaffe:1998hf}. 
The polarization of the incident beam is indicated with $\lambda_e$  and 
\begin{equation}
  A(y)=1-y+\frac{y^2}{2} \; , \quad B(y)=1-y \; , \quad C(y)=y(2-y) \; .
\label{eq:eqcoeffs}
\end{equation}
In Eq.~(\ref{eq:eqcross}), the indices $(\chi_1^{},\chi_1')$ refer to the chiralities 
of the entering quarks and identify each submatrix, while $(\chi_2^{},\chi_2')$ refer 
to the exiting quarks and point to the elements inside each submatrix. By expanding 
the sum over repeated indices in Eq.~(\ref{eq:crossh}), we get the expression
\bea
   \frac{d^7\sigma}{d\zeta\;dM_h^2\;d\phi_R\;dz\;dx\;dy\;d\phi_S} &=  &
   \sum\limits_{a} \; e_a^2 \, \frac{2 \alpha^2}{4\pi Q^2 y} \; \Big\{ A(y) \, f_1^a(x)
   \, D_1^a(z,\zeta,M_h^2) + \lambda_e \lambda \, \frac{C(y)}{2} \, g_1^a(x)\, D_1^a
   (z,\zeta,M_h^2)  \nn \\
   & & \quad + B(y) \,\frac{|\vec S_T| |\vec R_T|}{M_h}\, \sin (\phi_R+\phi_S)\, 
   h_1^a(x)\, H_1^{\open\,a}(z,\zeta,M_h^2) \Big\} \; .
\label{eq:crossh2}
\eea
For an unpolarized beam ($\lambda_e=0$, indicated with $O$) and a transversely polarized 
target ($\lambda=0$, indicated with $T$), Eq.~(\ref{eq:crossh2}) corresponds to Eq.~(10) 
of Ref.~\cite{Radici:2001na} after integrating over all transverse momenta. 
The following SSA can be built:
\bea
   A^{\sin (\phi_R+\phi_S)}_{OT}(y,x,z,M_h^2) &= &
   \frac{\int d\phi_S \,d\phi_R \,d\zeta \,\sin(\phi_R+\phi_S) \;d^7 \sigma_{OT}}
        {\int d\phi_S \,d\phi_R \,d\zeta \;d^7 \sigma_{OO}} \nn \\
    &= &|\vec S_T| \,\frac{B(y)}{A(y)} \; 
    \frac{\sum\limits_{a} \, e_a^2\,h_1^a(x)\,\int d\zeta \, \frac{|\vec R_T|}{2M_h} 
    \, H_1^{\open\,a}(z,\zeta,M_h^2)}
    {\sum\limits_{a} \, e_a^2\,f_1^a(x)\,\int d\zeta \, D_1^a(z,\zeta,M_h^2)} \; , 
\label{eq:ssa}
\eea
which allows to isolate the transversity $h_1$ at leading twist. Apart from the 
usual variables $x$, $y$, $z$, the only other variable to be measured is the angle 
$\phi_R + \phi_S$. Instead of using the scattering plane as a reference to measure 
azimuthal angles, it is sometimes convenient to use the directions of the beam and of 
the transverse component of the target spin. The new plane is rotated by the angle 
$\phi_S^{} \equiv -\phi_l^S$ with respect to the scattering plane; therefore, we have 
$\phi_R^{}\equiv\phi_R^S-\phi_l^S$ and 
$\phi_R^{}+\phi_S^{}\equiv\phi_R^S-2\phi_l^S$~\cite{Radici:2001na}.  

The asymmetry described in Eq.~(\ref{eq:ssa}) is the most general one at leading
twist for the case of two-hadron production when an unpolarized lepton beam scatters
off a transversely polarized target. No assumptions are made on the behavior of the
fragmentation functions. However, as we shall see in the next Section, it
is useful and desirable to understand how different partial waves
contribute to the above fragmentation functions.


\section{Partial-wave expansion for the two-hadron system}
\label{sec:LM}

If the invariant mass $M_h$ of the two hadrons is not very large, the pair can be
assumed to be produced mainly in the relative $s$-wave channel, with a typical 
smooth distribution, or in the $p$-wave channel with a Breit-Wigner
profile~\cite{Aubert:1983un}. Therefore, it is useful to expand 
Eq.~(\ref{eq:deltahtw}) -- or equivalently Eq.~(\ref{eq:deltactw}) -- in relative partial 
waves keeping only the first two harmonics. To this purpose, in the following we 
reformulate the kinematics in the cm frame of the two-hadron system. Then, the 
leading-twist projection for the quark-quark correlator $\Delta$ is conveniently expanded 
deducing a more detailed structure than Eq.~(\ref{eq:deltactw}). A set of new bounds is 
derived and the corresponding expression for the cross section is discussed.


\begin{figure}[h]
\includegraphics[width=6cm]{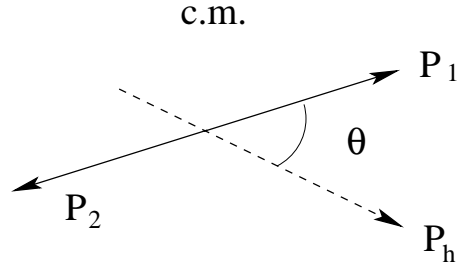} 
\caption {The hadron pair in the cm frame; $\theta$ is the cm polar angle of the pair with
respect to the direction of $P_h$ in the target rest frame.}
\label{fig:fig3}
\end{figure}


In the cm frame the emission of the two hadrons occurs  back-to-back. The direction 
identified by this emission forms an angle $\theta$ with the direction of $P_h$ in the
target rest frame (see Fig.~\ref{fig:fig3}). In this frame, the relevant variables become 
\bea
   P_h^\mu &= &\left[ \, \frac{M_h}{\sqrt{2}}, \, \frac{M_h}{\sqrt{2}}, \, 0, \, 0 
   \, \right] \nn \\
   R^\mu &= &\Bigg[ \, \frac{\sqrt{M_1^2+|\vec R|^2} - \sqrt{M_2^2+|\vec R|^2} -
   2 |\vec R| \cos\theta}{2\sqrt{2}} , \, 
   \frac{\sqrt{M_1^2+|\vec R|^2} - \sqrt{M_2^2+|\vec R|^2} +
   2 |\vec R| \cos\theta}{2\sqrt{2}} , \nn \\ 
   & &\qquad |\vec R| \sin\theta \cos\phi_R, \, |\vec R| \sin\theta \sin\phi_R \, 
   \Bigg] \nn \\
   \zeta &= &\frac{2R^-}{P_h^-} = \frac{1}{M_h} \, \left( \sqrt{M_1^2+|\vec R|^2} - 
   \sqrt{M_2^2+|\vec R|^2} - 2 |\vec R| \cos\theta \right) \; ,
\label{eq:kincm}
\eea
where 
\begin{equation}
   |\vec R| = \frac{1}{2M_h} \, \sqrt{M_h^2 - 2(M_1^2+M_2^2) + (M_1^2-M_2^2)^2} \; .
\label{eq:r}
\end{equation}
The crucial remark is that in this frame $\zeta$ depends linearly on
$\cos\theta$, i.e.\ $  \zeta = a + b \cos \theta \, ,$ with $a,b$, functions only of the 
invariant mass. This suggests that any function of $\zeta$ can be conveniently expanded 
in the basis of Legendre polynomials in $\cos\theta$, as discussed in the following.


\subsection{Partial-wave expansion of the quark-quark correlator $\Delta$ and positivity 
bounds}
\label{sec:deltalm}

We first express the leading-twist quark-quark correlator~(\ref{eq:deltahtw}) in terms 
of the cm variables. The connection between the two representations is defined as
\begin{equation} 
\Delta(z, \cos \theta, M_h^2, \phi_R) \equiv \frac{2 |\vec R|}{M_h}\,
	\Delta(z, \zeta, M_h^2, \phi_R),
\end{equation} 
to take into account the Jacobian of the transformation, 
$d \zeta = 2 |\vec R| / M_h \; d \cos \theta$. Therefore 
\begin{equation}
    {\mathcal P}_- \Delta_a(z,\cos\theta,M_h^2,\phi_R) \gamma^- 
    = \frac{2|\vec R| }{8 \pi M_h} 
   \, \left( D^a_1\big(z,\zeta(\cos\theta),M_h^2\big) + 
    i H_1^{\open\,a} \big(z,\zeta(\cos\theta),M_h^2\big) 
    \frac{|\vec R|}{M_h} \sin\theta \nslash_{\phi_R} \right) \, {\mathcal P}_- \; , 
\label{eq:deltacm}
\end{equation}
where $\nslash_{\phi_R} = \left[\, 0,\, 0,\, \cos\phi_R,\, \sin\phi_R\, \right]\; .$

The fragmentation functions can be expanded in Legendre polynomials as
\bea
   \frac{2 |\vec R|}{M_h}\,D_1\big(z,\zeta(\cos\theta),M_h^2\big) &= &\sum\limits_{n} 
   \, D_{1,n}(z,M_h^2) \, P_n (\cos\theta) \nn \\
   \frac{2 |\vec R|}{M_h}\,H_1^{\open}\big(z,\zeta(\cos\theta),M_h^2\big) &= 
   &\sum\limits_{n} \, H_{1,n}^{\open}(z,M_h^2) \, P_n (\cos\theta)  
\label{eq:fflm}
\eea
with
\bea 
   D_{1,n}(z,M_h^2) &= &\int_{-1}^{1} d\cos\theta \; P_n (\cos\theta) \; 
   \frac{2 |\vec R|}{M_h}\,D_1\big(z,\zeta(\cos\theta),M_h^2\big) \nn \\
   H_{1,n}^{\open}(z,M_h^2) &= &\int_{-1}^{1} d\cos\theta \; P_n (\cos\theta) \; 
   \frac{2 |\vec R|}{M_h}\,H_1^{\open}\big(z,\zeta(\cos\theta),M_h^2\big) \; .
\label{eq:fflmcoef}
\eea
We can truncate the expansion to the first three terms only $(n\leq 2)$, which are the 
minimal set required to describe all the ``polarization'' states of the system in the 
cm frame for relative partial waves with $L=0,1$. In fact, for $n=0$ $(P_0=1)$ the
correponding term in the correlator does not depend on $\theta$, it is 
``unpolarized''. For $n=1$, a 
term linear in $\cos\theta$ $(P_1=\cos\theta)$ describes the
interference between an ``unpolarized'' hadron pair in $s$-wave, for example on the 
left hand side of Fig.~\ref{fig:fig1}b, and a ``longitudinally polarized'' pair in
$p$-wave on the right hand side. Whenever in the correlator we encounter a term 
linear in $\sin\theta$, we will interpret it as the interference between a 
``unpolarized'' pair in $s$-wave and a ``transversely polarized'' pair in $p$-wave. 
Similarly, a term proportional to $\sin\theta \, \cos\theta$ indicates the 
interference between ``longitudinally'' and ``transversely polarized'' pairs always 
in a relative $p$-wave. The last case corresponds to $n=2$, that is interpreted as 
a ``tensor polarization'' still related to the intereference between pairs in a
relative $p$-wave. With notations that are consistent with previous
arguments, the correlator~(\ref{eq:deltacm}) is expanded as 
\bea
  {\mathcal P}_- \Delta(z,\zeta(\cos\theta),M_h^2,\phi_R) \gamma^- 
  &\sim &\frac{1}{8 \pi} \, \Big[ D_{1,0} (z,M_h^2) + D_{1,1}(z,M_h^2) \, \cos\theta + 
  D_{1,2}(z,M_h^2) \, \frac{1}{2} \, (3\cos^2\theta -1) \nn \\
  & &\quad + i\big( H_{1,0}^{\open}(z,M_h^2) + H_{1,1}^{\open}(z,M_h^2) \, \cos\theta 
  \big) \, \sin\theta \, \frac{|\vec R|}{M_h}\, \nslash_{\phi_R} \Big] \, 
  {\mathcal P}_- \nn \\
  &\equiv &\frac{1}{8 \pi} \, \Big[ D_{1,OO} (z,M_h^2) + D_{1,OL}(z,M_h^2) \, 
  \cos\theta + D_{1,LL}(z,M_h^2) \, \frac{1}{4} \, (3\cos^2\theta -1) \nn \\
  & &\quad + i\big( H_{1,OT}^{\open}(z,M_h^2) + H_{1,LT}^{\open}(z,M_h^2) \, 
  \cos\theta \big) \, \sin\theta \, \frac{|\vec R|}{M_h}\, \nslash_{\phi_R} \Big] \, 
  {\mathcal P}_- \; .
\label{eq:deltalm}
\eea

Consequently, the same correlator in the chiral basis becomes 
\bea
 [{\mathcal P}_- \Delta(z,\zeta,M_h^2,\phi_R) \gamma^-]_{\chi_2'\chi_2^{}} 
     \sim \hspace{9truecm}  \nn
\eea
\vspace{-.5truecm}
\begin{equation}
   \frac{1}{8 \pi} \, \left( \begin{array}{cc}
       \begin{array}{c} 
       D_{1,OO} (z,M_h^2) + D_{1,OL}(z,M_h^2) \, \cos\theta \\
       + D_{1,LL}(z,M_h^2) \, \frac{1}{4} \, (3\cos^2\theta -1) \end{array} & 
       \begin{array}{c}
       i e^{i\phi_R} \frac{|\vec R|}{M_h} \, \sin\theta \hspace{3truecm} \\
       \times \big( H_{1,OT}^{\open}(z,M_h^2) + H_{1,LT}^{\open}(z,M_h^2) \, 
       \cos\theta \big) \end{array} 
       \\[20pt]
       \begin{array}{c}
       -i e^{-i\phi_R} \frac{|\vec R|}{M_h} \, \sin\theta \hspace{3truecm}\\
       \times \big( H_{1,OT}^{\open}(z,M_h^2) + H_{1,LT}^{\open}(z,M_h^2) \, 
       \cos\theta \big) \end{array} &
       \begin{array}{c} 
       D_{1,OO} (z,M_h^2) + D_{1,OL}(z,M_h^2) \, \cos\theta \\
       + D_{1,LL}(z,M_h^2) \, \frac{1}{4} \, (3\cos^2\theta -1) \end{array} 
       \end{array} \right) \; . 
\label{eq:deltalmc}
\end{equation}
It is useful to project out of Eq.~(\ref{eq:deltalmc}) the information about the 
orbital angular momentum of the system, which is encoded in the angular distribution 
of the hadron pair. In fact, for $L\leq1$ the decay matrix for the hadron pair is 
given by the following bilinear combination of spherical harmonics:
\bea
  {\mathcal D}_{MM'}^{LL'}(\theta,\phi_R) = Y_{LM} \, Y^*_{L'M'} = 
  \hspace{10truecm} \nn 
\eea
\vspace{-.5truecm}
\begin{equation}
   \frac{1}{4 \pi}\left( \begin{array}{c|ccc}
    1 & -\sqrt{\frac{3}{2}} \, \sin\theta \, e^{i\phi_R} & \sqrt{3}\cos\theta & 
    \sqrt{\frac{3}{2}}\, \sin\theta \, e^{-i\phi_R} \\[7pt]
    \hline \\[-12pt]
    -\sqrt{\frac{3}{2}}\, \sin\theta \, e^{-i\phi_R} & \frac{3}{2}\,\sin^2\theta & 
     -\frac{3}{\sqrt{2}}\, \cos\theta\sin\theta\,e^{-i\phi_R} & 
     -\frac{3}{2}\,\sin^2\theta\,e^{-2i\phi_R} \\[2pt]
    \sqrt{3}\cos\theta & -\frac{3}{\sqrt{2}}\, \cos\theta\sin\theta\,e^{i\phi_R} & 
     3\,\cos^2\theta & \frac{3}{\sqrt{2}}\, \cos\theta\sin\theta\,e^{-i\phi_R} 
     \\[2pt]
    \sqrt{\frac{3}{2}} \, \sin\theta \, e^{i\phi_R} & 
      -\frac{3}{2}\,\sin^2\theta\,e^{2i\phi_R} & 
      \frac{3}{\sqrt{2}}\, \cos\theta\sin\theta\,e^{i\phi_R} & 
      \frac{3}{2}\,\sin^2\theta \end{array} \right) \; ,
\label{eq:decay}
\end{equation}
with $L,L'\leq1$ and $|M^{\left(\prime\right)}|\leq
L^{\left(\prime\right)}$. The upper left block corresponds to $L=L'=0$,
i.e.\ to the system being in relative $s$ wave. The lower right block instead
corresponds to $L=L'=1$, i.e.\ to the system being in relative $p$ wave,
including all the contributions corresponding to different $M,M'$ projections and their 
interferences. The off-diagonal blocks indicate, obviously, the interference
between the $s$ and $p$ waves. Using the decay matrix, it is possible to represent
the fragmentation in the basis of the quark chirality and of the pair orbital
angular momentum. In fact, Eq.~(\ref{eq:deltalmc}) can be written as
\begin{equation}
 [{\mathcal P}_-\Delta(z,\zeta,M_h^2,\phi_R) \gamma^-]_{\chi_2'\chi_2^{}} 
 = [{\mathcal P}_-\Delta(z,M_h^2) \gamma^-]_{M'M\,\chi_2'\chi_2^{}}^{L'L} 
  \; {\mathcal D}_{MM'}^{LL'}(\theta,\phi_R) \; ,
\label{eq:deltalmc2}
\end{equation}
where
\begin{equation} 
 [{\mathcal P}_- \Delta(z,M_h^2) \gamma^-]_{M'M\,\chi_2'\chi_2^{}}^{L'L} 
 = \frac{1}{8} \; \left( \begin{array}{cc}
     A_{M'M}^{L'L} & B_{M'M}^{L'L} \\[2pt]
     (B_{M'M}^{L'L})^{\dagger} & A_{M'M}^{L'L} \end{array} \right) 
\label{eq:deltalmc3}
\end{equation}
and
\bea
   A_{M'M}^{L'L} &= &\left( \begin{array}{c|ccc}
       D_{1,OO}^s & 0 & \frac{2}{\sqrt{3}} \, D_{1,OL} & 0 \\[7pt]
       \hline \\[-12pt]
       0 & D_{1,OO}^p-\frac{1}{3}\,D_{1,LL} & 0 & 0 \\[2pt]
       \frac{2}{\sqrt{3}}\,D_{1,OL} & 0 & D_{1,OO}^p+\frac{2}{3}\,D_{1,LL} & 0 \\[2pt]
       0 & 0 & 0 & D_{1,OO}^p - \frac{1}{3}\,D_{1,LL} \end{array} \right)  ,
\label{eq:almc} \\[10pt]
   B_{M'M}^{L'L} &= &\left( \begin{array}{c|ccc} 
      0 & 0 & 0 & i \frac{2\sqrt{2}}{\sqrt{3}} \, \frac{|\vec R|}{M_h}\,
      H_{1,OT}^{\open} \\[7pt]
      \hline \\[-12pt]
      -i \frac{2\sqrt{2}}{\sqrt{3}} \, \frac{|\vec R|}{M_h}\, H_{1,OT}^{\open} & 0 & 
      -i \frac{2\sqrt{2}}{3} \, \frac{|\vec R|}{M_h}\, H_{1,LT}^{\open} & 0
      \\[2pt]
      0 & 0 & 0 & i \frac{2\sqrt{2}}{3} \, \frac{|\vec R|}{M_h}\, 
      H_{1,LT}^{\open} \\[2pt]
      0 & 0 & 0 & 0 \end{array} \right). 
\label{eq:blmc}
\eea
The fragmentation matrix 
$[{\mathcal P}_- \Delta(z,M_h^2) \gamma^-]_{M'M\,\chi_2'\chi_2^{}}^{L'L}$
fulfills all the fundamental properties, namely Hermiticity, parity 
invariance~\cite{Jaffe:1998pv} and angular momentum conservation 
$(\chi_2'+M=\chi_2^{}+M')$. The imaginary entries in its off-diagonal submatrix are 
T-odd fragmentation functions. It is worth noticing that with the 
projection~(\ref{eq:deltalmc2}) we gained a further information on the ``unpolarized'' 
state of the hadron pair. In fact, we see from the diagonal of Eq.~(\ref{eq:almc}) 
that the spherically symmetric state in the cm frame receives contributions from both 
the relative $s$ and $p$ waves, such that when performing the matrix multiplication of 
Eq.~(\ref{eq:deltalmc2}) we get
\begin{equation} 
   D_{1,OO}(z,M_h^2) = \frac{1}{4}\,D_{1,OO}^s(z,M_h^2) + \frac{3}{4}\, 
   D_{1,OO}^p(z,M_h^2) \; .
\label{eq:dOO}
\end{equation}
However, in an actual cross section the two contributions are merged together and are
kinematically indistinguishable, unless a specific hypothesis on the dependence upon
the invariant mass $M_h$ is assumed for the two different partial waves, e.g. a
resonant contribution for the $p$ wave and a continuum background for the $s$ wave.

Finally, from the matrix~(\ref{eq:deltalmc3}) being positive semidefinite the 
following bounds are derived:\footnote{Note that the bounds involving the pure p-wave 
functions correspond to those obtained in Ref.~\cite{Bacchetta:2001rb}}
\bea
   D_{1,OO}^s &\geq &0, \qquad D_{1,OO}^p \geq 0, \nn \\
   -\frac{3}{2}\,D_{1,OO}^p &\leq &D_{1,LL} \leq 3 D_{1,OO}^p, \nn \\
   D_{1,OL} &\leq &\sqrt{\frac{3}{4}\,D_{1,OO}^s \left(D_{1,OO}^p+\frac{2}{3}\,
   D_{1,LL}\right)} \leq \frac{3}{2}\,D_{1,OO}, \nn\\
   \frac{|\vec R|}{M_h}\,H_{1,OT}^{\open} &\leq &\sqrt{\frac{3}{8}\,D_{1,OO}^s
   \left(D_{1,OO}^p-\frac{1}{3}\,D_{1,LL}\right)} \leq \frac{3}{2}\,D_{1,OO}, \nn\\
   \frac{|\vec R|}{M_h}\,H_{1,LT}^{\open} &\leq
   &\frac{3}{2\sqrt{2}}\,\sqrt{\left(D_{1,OO}^p+\frac{2}{3}\,D_{1,LL}\right)\,
   \left(D_{1,OO}^p-\frac{1}{3}\,D_{1,LL}\right)}\leq \frac{9}{8}\,D_{1,OO} \; .
\label{eq:lmbound}
\eea


\subsection{Cross section and transverse spin asymmetries}
\label{sec:crosslm}

Using Eq.~(\ref{eq:deltalmc2}) inside Eq.~(\ref{eq:crossh}), we can take advantage of
the full power of the analysis in the helicity formalism. In fact, the cross section
can be expanded in the density matrices for the target helicity, for the chirality 
of the initial and fragmenting quark, and for the relative orbital angular momentum 
of the leading hadron pair~\cite{Jaffe:1998hf}. Inserting the corresponding 
expressions~(\ref{eq:rho},\ref{eq:phihctw},\ref{eq:eqcross},\ref{eq:deltalmc3},\ref{eq:decay}), 
we get   
\begin{eqnarray}
\lefteqn{\displaystyle{\frac{d^7\sigma}{d\zeta\;dM_h^2\;d\phi_R\;dz\;dx\;dy\;d\phi_S}}}
\nn \\
&=&   
 \sum\limits_{a} \; \rho_{\Lambda\Lambda'}(S) \, 
 [{\mathcal P}_+ \Phi_a(x) \gamma^+]_{\chi_1'\chi_1^{}}^{\Lambda'\Lambda}
 \; \left( \displaystyle{\frac{d\sigma^{eq_a}}{dy}} 
   \right)^{\chi_1^{}\chi_1'\,;\,\chi_2^{}\chi_2'} \; 
 [{\mathcal P}_- \Delta(z,M_h^2) \gamma^-]_{M'M\,\chi_2'\chi_2^{}}^{L'L} 
   \; {\mathcal D}_{MM'}^{LL'}(\theta,\phi_R)  \nn \\
 & =& \sum\limits_{a} \; e_a^2 \, \displaystyle{\frac{\alpha^2}{2 \pi Q^2 y}} \; 
 \biggl\{ \biggl[ A(y) \, f_1^a(x) + \lambda_e \lambda \, 
 \displaystyle{\frac{C(y)}{2}} \, g_1^a(x) \biggr] \, \biggl[ 
 \displaystyle{\frac{D_{1,OO}^s+3D_{1,OO}^p}{4}} + D_{1,OL} \cos\theta + D_{1,LL} 
 \displaystyle{\frac{1}{4}}\, (3\cos^2\theta -1)\biggr]\nn  \\
&&   + B(y) \,\displaystyle{\frac{|\vec S_T| |\vec R|}{M_h}}\, \sin (\phi_R+\phi_S)\, 
   h_1^a(x)\, \sin\theta \, \Big[ H_{1,OT}^{\open}+ H_{1,LT}^{\open} \cos\theta \Big] 
   \biggr\} \; , 
\label{eq:crossh3}
\end{eqnarray}
where all the fragmentation functions depend just on $(z,M_h^2)$. 

Replacing $\lambda=\lambda_e=|\vec S_T|=0$ in the previous equation, we get the 
unpolarized cross section $d^7\sigma_{OO}$. However, it is particularly interesting to 
consider the case for an unpolarized beam and a transversely polarized target, i.e.\ 
\begin{equation}
  \frac{d^7\sigma_{OT}}{d\zeta\;dM_h^2\;d\phi_R\;dz\;dx\;dy\;d\phi_S} = 
  \sum\limits_{a} \; e_a^2 \, \frac{\alpha^2}{2 \pi Q^2 y} \; B(y) \,
  \frac{|\vec S_T| |\vec R|}{M_h}\, \sin (\phi_R+\phi_S)\, 
   h_1^a(x)\, \sin\theta \, \Big[ H_{1,OT}^{\open}+ H_{1,LT}^{\open} \cos\theta \Big] 
   \; ,
\label{eq:crosshlmOT}
\end{equation}
because we can see that the transversity $h_1$ can be matched by two different
chiral-odd, T-odd IFF: one ($H_{1,OT}^{\open}$) pertaining to the interference between 
$s$- and $p$-wave states of the hadron pair, the other ($H_{1,LT}^{\open}$) pertaining
to the $p$ wave only. The partial-wave analysis allows us for the first time to 
comprehend different theoretical analyses in a unifying framework. In fact, 
$H_{1,OT}^{\open}$ corresponds to the hypothesis first formulated in 
Ref.~\cite{Jaffe:1998hf}, and later reconsidered in Ref.~\cite{Radici:2001na}, where 
the necessary spin asymmetry is generated by the interference between two channels 
describing two leading pions in the relative $s$ and $p$ waves, respectively. As a 
simple cross-check, taking Eq.~(\ref{eq:crosshlmOT}) and integrating the $\theta$ 
dependence away yields
\bea
   \int_{-1}^1 d\cos\theta\, 
   \frac{d^7\sigma_{OT}}{d\cos\theta\;dM_h^2\;d\phi_R\;d\phi_S\;dz\;dx\;dy} &= &
   \int_{-1}^1 d\cos\theta\; 
   \frac{d^7\sigma_{OT}}{d\zeta\;dM_h^2\;d\phi_R\;d\phi_S\;dz\;dx\;dy} \nn \\[2pt]
   &= &\sum\limits_{a} \; e_a^2 \, \frac{\alpha^2}{4Q^2 y} \; 
   B(y) \,\frac{|\vec S_T| |\vec R|}{M_h}\, \sin(\phi_R+\phi_S)\, h_1^a(x)\, 
   H_{1,OT}^{\open}(z,M_h^2) \; .
\label{eq:jaffecmp}
\eea
This asymmetry corresponds to the one studied in Ref.~\cite{Jaffe:1998hf}, although 
in that paper several assumptions were made. Firstly, the IFF was factorized in a 
part dependent only on the variable $z$, designated as $\delta {\hat q}_I(z)$, and in a 
part containing the $M_h$-dependent $\pi$-$\pi$ phase shifts, 
$\sqrt{6}\sin\delta_0 \sin\delta_1 \sin(\delta_0-\delta_1)$. Secondly, the azimuthal 
angle of the target spin was taken $\phi_S=0$, due to neglecting the scattering 
angle (see Fig.~\ref{fig:fig2}). The azimuthal angle of the hadron pair defined in 
Ref.~\cite{Jaffe:1998hf} is $\phi=\pi/2-\phi_R$. It is worth to note that the peculiar 
behavior in the invariant mass discussed in Ref.~\cite{Jaffe:1998hf} relies on the 
assumption that only the $\pi$-$\pi$ rescattering can generate the T-odd character of the 
IFF. It has been already shown, however, that a different model with more general 
assumptions leads to a unfactorized $(z,M_h^2)$ dependence of the 
fragmentation function and to a completely different behaviour of the 
SSA~\cite{Radici:2001na}. Therefore, it is of great interest to experimentally explore the
production of two unpolarized hadrons, e.g. two pions, in the relevant kinematic range,
namely with an invariant mass around the $\rho$ resonance. 

As for the function $H_{1,LT}^{\open}$, it naturally links with the analysis developed in 
the case of a spin-1 hadron fragmentation~\cite{Bacchetta:2000jk}, because the two 
spinless hadrons, e.g., two pions, can be considered as the decay product of a spin-1 
resonance, e.g., a $\rho$ particle. The T-odd IFF arise from the interference between two 
different channels in the relative $p$ wave. To the purpose of isolating an asymmetry 
containing the function $H_{1,LT}^{\open}$, we show that integrating 
Eq.~(\ref{eq:crosshlmOT}) upon $\theta$ in a different range, namely in the interval 
$[-\pi/2,\pi/2]$, yields
\bea
   \int_{-\frac{\pi}{2}}^{\frac{\pi}{2}} d\theta &\sin\theta\, &
   \frac{d^7\sigma_{OT}}{d\cos\theta\;dM_h^2\;d\phi_R\;dz\;dx\;dy\;d\phi_S} = 
   \int_{-\frac{\pi}{2}}^{\frac{\pi}{2}} d\theta\sin\theta\; 
   \frac{d^7\sigma_{OT}}{d\zeta\;dM_h^2\;d\phi_R\;dz\;dx\;dy\;d\phi_S}
   \hspace{2truecm} \nn \\[2pt]
   &= &\sum\limits_{a} \; e_a^2 \, \frac{\alpha^2}{4Q^2 y} \;  
   B(y) \,\frac{|\vec S_T| |\vec R|}{M_h}\, \sin(\phi_R+\phi_S)\, h_1^a(x)\, 
   \Big[\, H_{1,OT}^{\open}(z,M_h^2) +
   \frac{4}{3 \pi}\,H_{1,LT}^{\open}(z,M_h^2) \Big]  \; ,
\label{eq:ppterm}
\eea
where both kinds of IFF appear at leading twist and can contribute to a SSA isolating
the transversity $h_1$. Although spin-1 fragmentation functions have already been 
proposed in the past as possible chiral-odd partners for the 
transversity~\cite{Efremov:1982sh,Ji:1994vw,Anselmino:1996vq,Bacchetta:2000jk}, to our 
knowledge this is the first time that the asymmetry where they occur is explicitly 
identified and a clear distinction from the $s$-$p$ interference is made.
 
There are not yet quantitative model predictions for $H_{1,LT}^{\open}$; on the other 
hand, since the $p$-wave production of two hadrons becomes significant only when it 
proceeds via a spin-1 resonance, we can expect that the shape of this function in the
invariant mass corresponds to a Breit-Wigner curve peaked at the resonance mass.
Moreover, it has the same features as a single-particle fragmentation function, 
unlike $H_{1,OT}^{\open}$: its evolution equations can be expected to be analogous to 
that of the transversity~\cite{Boer:2001zr,Stratmann:2001pt,Boer:2001zw}; 
it does not require a rescattering of the hadrons after they are produced
and its physical origin could have something in common with the one of the 
Collins function. However, it should be noticed that in the case of the Collins 
function an essential role is played by the partonic transverse momentum, which in the 
case of $H_{1,LT}^{\open}$ is replaced  by the relative transverse momentum of the 
hadron pair.

It would be interesting to elaborate on these topics since data for the 
electromagnetic $\rho$ production and decay are already available in the diffractive 
regime~\cite{Breitweg:1999fm,Ackerstaff:2000bz,Adloff:1999kg}, and they could be 
available in the DIS regime as well in the near future.


\section{Explicit dependence on the transverse momenta}
\label{sec:LM-T}

For sake of completeness, in this Section we extend the previous results to the 
case where the transverse momenta are not integrated away. In this case, the cross 
section is nine-fold and reads
\be
   \frac{d^9\sigma}{d\zeta\;dM_h^2\;d\phi_R\;dz\;d\vec P_{h \perp}\;dx\;dy\;d\phi_S} = 
   \sum\limits_{a}\; \frac{\alpha^2 \, y\, e_a^2}{32\,z\, Q^4} \; L_{\mu\nu} \; 
   2M \, W_a^{\mu \nu} \; .
\label{eq:crossdiff}
\ee 
The hadronic tensor takes the form
\begin{equation} 
    2M W_a^{\mu\nu} = 32z \; {\mathcal I} \Big[ \, \mbox{Tr} \big[ 
    \Phi_a(x,\vec p_T,S) \, \gamma^\mu \, \Delta_a(z,\vec k_T,\zeta,M_h^2,\phi_R) \, 
    \gamma^\nu \big] \, \Big] \; ,
\label{eq:wconv2}
\end{equation}
where we introduced the shorthand notation
\begin{equation} 
    {\mathcal I} [f] \equiv \int d\vec p_T d\vec k_T \, 
    \delta (\vec p_T - \vec P_{h \perp}/z - \vec k_T) \, [f] \; ,
\label{eq:conv}
\end{equation} 
and where the transverse momentum dependent correlation functions are
\begin{eqnarray} 
    \Phi_a (x,\vec p_T,S) &= &\int d p^-\,\Phi_a(p;P,S) \Big|_{p^+ = x P^+} \; ,
\label{eq:phipt} \\
    \Delta_a (z,\vec k_T,\zeta,M_h^2,\phi_R) &= &\frac{1}{32z} \; 
    \int d k^+\,\Delta_a(k;P_h,R) \Big|_{k^- = P_h^-/z} \; . 
\label{eq:deltakt}
\end{eqnarray} 

The leading-twist projection of $W^{\mu\nu}$ proceeds in an analogous way to 
Eq.~(\ref{eq:wh}); we usually have~\cite{Mulders:1996dh}
\bea
    {\mathcal P}_+ \Phi_a(x,\vec p_T,S) \gamma^+ &= &\Big\{ 
    f_1^a(x,\vec p_T^{\,2}) + \frac{\epsilon_{T\rho\sigma} \,S_T^\rho p_T^\sigma}{M} 
    \,f_{1T}^{\perp\,a}(x,\vec p_T^{\,2}) + i h_1^{\perp\,a}(x,\vec p_T^{\,2})
    \frac{\pslash_T}{M} \nn \\
    & &+ \big[\lambda g_{1L}^a(x,\vec p_T^{\,2}) + \frac{\vec p_T\cdot \vec
    S_T}{M}\,g_{1T}^a(x,\vec p_T^{\,2}) \big] \gamma_5 \nn \\
    & &+ \big[\lambda h_{1L}^{\perp\,a}(x,\vec p_T^{\,2}) + \frac{\vec p_T\cdot \vec
    S_T}{M}\,h_{1T}^{\perp\,a}(x,\vec p_T^{\,2}) \big] \gamma_5 \frac{\pslash_T}{M} 
    + h_{1T}^a(x,\vec p_T^{\,2}) \,\gamma_5\,\Sslash_T \Big\} \; ,
\label{eq:phihtw-T}
\eea
where $\epsilon_T^{\mu\nu} = \epsilon^{-+\mu\nu}$. Equation~(\ref{eq:phihtw-T})
corresponds to Eq.(2) of Ref.~\cite{Radici:2001na}. Again, similarly to 
Eq.~(\ref{eq:wctw2}) and following ones, we project out the density matrix of the 
target helicity so that Eq.~(\ref{eq:phihtw-T}) in the basis of quark chirality and 
target helicity becomes
\bea
    [{\mathcal P}_+ \Phi_a \gamma^+]_{\chi_1'\chi_1^{}}^{\Lambda'\Lambda} = 
    \hspace{11truecm} \nn
\eea
\vspace{-.5truecm}
\begin{equation}
    =  \left( \begin{array}{cc|cc}
              f_1^a+g_{1L}^a & \frac{|\vec p_T|}{M}\,e^{-i\phi_p}\,\lf(g_{1T}^{a}+ 
	      i f_{1T}^{\perp\,a}\rg) & 
	      \frac{|\vec p_T|}{M}\,e^{-i\phi_p}\,\lf(h_{1L}^{\perp\,a}+
	      i h_{1}^{\perp\,a}\rg) & 
	       \frac{|\vec p_T|^2}{M^2}\,e^{-2i\phi_p}\,h_{1T}^{\perp\,a} \\[2pt]
	      \frac{|\vec p_T|}{M}\,e^{i\phi_p}\,\lf(g_{1T}^{a}- 
	      i f_{1T}^{\perp\,a}\rg) & f_1^a - g_{1L}^a & 
	      2h_1^a & -\frac{|\vec p_T|}{M}\,e^{-i\phi_p}\,\lf(h_{1L}^{\perp\,a}-
	      i h_{1}^{\perp\,a}\rg) \\[7pt]
	      \hline \\[-12pt]
	      \frac{|\vec p_T|}{M}\,e^{i\phi_p}\,\lf(h_{1L}^{\perp\,a}-
	      i h_{1}^{\perp\,a}\rg) & 2h_1^a & 
	      f_1^a - g_{1L}^a & -\frac{|\vec p_T|}{M}\,e^{-i\phi_p}\,\lf(g_{1T}^{a}- 
	      i f_{1T}^{\perp\,a}\rg) \\[2pt]
	      \frac{|\vec p_T|^2}{M^2}\,e^{2i\phi_p}\,h_{1T}^{\perp\,a} & 
	      -\frac{|\vec p_T|}{M}\,e^{i\phi_p}\,\lf(h_{1L}^{\perp\,a}+
	      i h_{1}^{\perp\,a}\rg)& 
	       -\frac{|\vec p_T|}{M}\,e^{i\phi_p}\,\lf(g_{1T}^{a}+ 
	      i f_{1T}^{\perp\,a}\rg) & f_1^a + g_{1L}^a \end{array} \right) \; , 
\label{eq:phihctw-T}
\end{equation}
where $\phi_p$ is the azimuthal angle of $\vec p_T$. The matrix is Hermitean,
respects parity invariance and conservation of total angular momentum. Introducing the 
dependence upon the quark transverse momentum $\vec p_T$ modifies the conditions
for angular momentum and parity conservation, which now read, respectively,
\bea
\Lambda'_{1}+\chi'_{1}+l_{p_T} &= &\Lambda_{1} +\chi_{1} \nn \\
\left[ {\cal P}_+ \Phi \g^+ \right]_{\chi_1' \chi_1^{}}^{\Lambda'\Lambda} &=
&(-1)^{l_{p_T}} \, \left[ {\cal P}_+ \Phi \g^+ \right]_{-\chi^{}_{1}\, -\chi'_{1}}
^{-\Lambda'\, -\Lambda} \biggr \rvert_{l_{p_T}\rightarrow - l_{p_T}} \; ,
\eea
where $l_{p_T}$ denotes the units of angular momentum introduced by $\vec p_T$. 
The chiral transposed matrix is still positive definite, so that the bounds on the 
various distribution functions can be obtained~\cite{Bacchetta:1999kz}.

The leading-twist projection of the fragmenting quark correlator is
\bea
   {\mathcal P}_- \Delta_a(z,\vec k_T,\zeta,M_h^2,\phi_R) \gamma^- &= &
    \frac{1}{8 \pi} \, \left( D^a_1(z,\zeta,M_h^2,\vec k_T^2,\vec k_T\cdot\vec R_T) + 
    i {\bar H}_1^{\open\,a } (z,\zeta,M_h^2,\vec k_T^2,\vec k_T\cdot\vec R_T) 
    \frac{\Rslash_T}{M_h} \right. \nn \\
    & &\hspace{-2.5truecm} \left. + 
    i H_1^{\perp\,a} (z,\zeta,M_h^2,\vec k_T^2,\vec k_T\cdot\vec R_T) 
    \frac{\kslash_T}{M_h} + 
    G_1^{\perp\,a} (z,\zeta,M_h^2,\vec k_T^2,\vec k_T\cdot\vec R_T) \, 
    \frac{\epsilon_T^{\mu\nu} R_{T\,\mu} k_{T\nu}}{M_h^2}\, \gamma_5 \right) \, 
    {\mathcal P}_- \; ,
\label{eq:deltahtw-T}
\eea
where the actual dependence of the fragmentation functions is the most general one
possible~\cite{Bianconi:1999cd}. In Eq.~(\ref{eq:deltahtw-T}) 
${\mathcal P}_- \Delta_a$ corresponds to Eq.(3) of Ref.~\cite{Radici:2001na}. New 
functions appear: $G_1^{\perp}$ is chiral even but T-odd, $H_1^{\perp}$ is chiral odd 
and T-odd and represents the analogue of the Collins effect for a two-hadron
emission~\cite{Bianconi:1999cd}. Upon integration over $d\vec k_T$, $G_1^{\perp}$ 
vanishes and the surviving parts of ${\bar H}_1^{\open}$ and $H_1^{\perp}$ merge into 
the function $H_1^{\open}$ of Eq.~(\ref{eq:deltahtw}) keeping the same $\Rslash_T/M_h$ 
structure. In the chiral basis of the fragmenting quark, Eq.~(\ref{eq:deltahtw-T}) 
becomes
\begin{equation}
 [{\mathcal P}_- \Delta_a(z,\vec k_T,\zeta,M_h^2,\phi_R) \gamma^-]_{\chi_2'\chi_2^{}} 
 = \frac{1}{8 \pi} \left( \begin{array}{cc}
       D_1^a + \frac{\epsilon_T^{\mu\nu} R_{T\,\mu} k_{T\nu}}{M_h^2}\,G_1^{\perp\,a} 
       & i \Big( e^{i\phi_R} \frac{|\vec R_T|}{M_h} \, {\bar H}_1^{\open\,a} 
       + e^{i\phi_k} \frac{|\vec k_T|}{M_h} \, H_1^{\perp\,a} \Big) \\[5pt]
       -i \Big( e^{-i\phi_R} \frac{|\vec R_T|}{M_h} {\bar H}_1^{\open\,a} 
        + e^{-i\phi_k} \frac{|\vec k_T|}{M_h} \, H_1^{\perp\,a} \Big) & 
       D_1^a - \frac{\epsilon_T^{\mu\nu} R_{T\,\mu} k_{T\nu}}{M_h^2}\,G_1^{\perp\,a} 
      \end{array} \right) \; .
\label{eq:deltactw-T}
\end{equation}
The following bounds are derived:
\bea
   \frac{|\epsilon_T^{\mu\nu} R_{T\mu} k_{T\nu}|}{M_h^2}\,|G_1^{\perp\,a}| &\leq 
   &D_1^a \nn \\
   \frac{|\vec R_T|^2}{M_h^2} \, ({\bar H}_1^{\open\,a})^2 + 
   \frac{|\vec k_T|^2}{M_h^2} \, (H_1^{\perp\,a})^2 + 
   \frac{2\vec k_T\cdot \vec R_T}{M_h^2} \, {\bar H}_1^{\open\,a}\, H_1^{\perp\,a}
   &\leq &(D_1^a)^2- 
   \frac{|\epsilon_T^{\mu\nu} R_{T\mu} k_{T\nu}|^2}{M_h^4}\,|G_1^{\perp\,a}|^2 \; .
\label{eq:ffbound-T}
\eea

Expanding the cross section of Eq.~(\ref{eq:crossdiff}) along the same lines leading
to Eq.~(\ref{eq:crossh}), we have
\bea
\frac{d^9\sigma}{d\zeta\;dM_h^2\;d\phi_R\;d\vec P_{h \perp}\;dz\;dx\;dy\;d\phi_S} &= &
\sum\limits_{a} \; \rho_{\Lambda\Lambda'}(S) \, {\cal I}\;\biggl[
[{\mathcal P}_+ \Phi_a(x,\vec p_T) \g^+]_{\chi_1'\chi_1^{}}^{\Lambda'\Lambda}
\; \left( \frac{d\sigma^{eq_a}}{dy} \right)^{\chi_1^{}\chi_1'\,;\,\chi_2^{}\chi_2'} 
\nn \\
& &\qquad \times [{\mathcal P}_- \Delta_a(z,\vec k_T,\zeta,M_h^2,\phi_R) \g^-]_{\chi_2' 
\chi_2^{}}\biggr]  \; , 
\label{eq:crossdiffh}
\eea
where $[{\mathcal P}_+ \Phi_a\gamma^+]$ and $[{\mathcal P}_- \Delta_a\gamma^-]$ are 
given by Eqs.~(\ref{eq:phihctw-T}) and (\ref{eq:deltactw-T}), respectively. The 
complete formula for the cross section is given in App.~\ref{a:cswithT}.


\section{Partial-wave expansion with transverse momenta}
\label{sec:LM-T2}

It is again useful to expand all the fragmentation functions of
Eq.~(\ref{eq:deltahtw-T}) in the relative partial waves of the hadron pair. The
dependence on $\vec k_T\cdot \vec R_T$ makes the expansion more involved:
\bea
   D_1 &= &D_{1,OO} + D_{1,OL}\cos\theta + D_{1,LL} \frac{1}{4}\,(3\cos^2\theta -1) +
   \cos(\phi_k-\phi_R)\, \sin\theta\,(D_{1,OT} + D_{1,LT} \cos\theta) 
   \nn \\
   & &\qquad + \cos(2\phi_k-2\phi_R)\, \sin^2\theta \, D_{1,TT} \nn \\[4pt]
   G_1^\perp &= &G_{1,OT}^\perp + G_{1,LT}^\perp \cos\theta + \cos(\phi_k-\phi_R)\,
   \sin\theta\, G_{1,TT}^\perp \nn \\[4pt]
   \bar{H}_1^{\open} &= &\bar{H}_{1,OT}^{\open} + \bar{H}_{1,LT}^{\open} \cos\theta + 2
   \cos(2\phi_k-2\phi_R)\, \sin\theta \, \bar{H}_{1,TT}^{\open}\nn \\
   H_1^\perp &= &H_{1,OO}^\perp + H_{1,OL}^\perp\cos\theta + H_{1,LL}^\perp 
   \frac{1}{4}\,(3\cos^2\theta -1) + 2\cos(\phi_k-\phi_R)\, \sin\theta\,
   (H_{1,OT}^\perp + H_{1,LT}^\perp \cos\theta) \nn \\
   & &\qquad + 2\cos(2\phi_k-2\phi_R)\, \sin^2\theta \, H_{1,TT}^\perp 
	-\sin^2 \theta \frac{|\vec{R}|}{|\vec{k}_T|}\, H_{1,TT}^{\open} \; , 
\label{eq:fflm-T}
\eea
where all the functions depend on $(z,\vec k_T^2,M_h^2)$. Then, similarly to
Eq.~(\ref{eq:deltalmc2}), Eq.~(\ref{eq:deltactw-T}) can be further expanded in the
basis of the pair orbital angular momentum as
\begin{equation}
 [{\mathcal P}_- \Delta(z,\vec k_T,\zeta,M_h^2,\phi_R) \gamma^-]_{\chi_2'\chi_2^{}} = 
  [{\mathcal P}_- \Delta(z,\vec k_T^2,M_h^2) \gamma^-]_{M'M\,\chi_2'\chi_2^{}}^{L'L} \;
  {\mathcal D}_{MM'}^{LL'}(\theta,\phi_k,\phi_R) \; .
\label{eq:deltalmc2-T}
\end{equation}
The full expression of 
$[{\mathcal P}_- \Delta(z,\vec k_T^2,M_h^2) \gamma^-]$ is shown in App.~\ref{a:app2}. 
The fully expanded differential cross section in the helicity basis of target, initial 
and final quark, as well as in the basis of orbital angular momentum of the hadron
pair is then
\bea
 \frac{d^9\sigma}{d\zeta\;dM_h^2\;d\phi_R\;d\vec P_{h \perp}\;dz\;dx\;dy\;d\phi_S} &= &
 \sum\limits_{a} \; \rho_{\Lambda\Lambda'}(S) \, {\cal I}\;\biggl[
 [{\mathcal P}_+ \Phi_a(x,\vec p_T) \g^+]_{\chi_1'\chi_1^{}}^{\Lambda'\Lambda}
 \; \left( \frac{d\sigma^{eq_a}}{dy} \right)^{\chi_1^{}\chi_1'\,;\,\chi_2^{}\chi_2'} 
 \nn \\
& &\qquad \times [{\mathcal P}_- \Delta_a(z,\vec k_T,\zeta,M_h^2,\phi_R) 
   \g^-]_{M'M\,\chi_2'\chi_2^{}}^{L'L}\biggr] \;{\mathcal D}_{MM'}^{LL'}(\theta,\phi_R). 
\label{eq:crossdiffhpw}
\eea
Its explicit expression is presented in App.~\ref{a:sp}. The pure $p$-wave sector 
corresponds to the cross section for the production of a polarized spin-1 hadron and has 
already been fully studied in Refs.~\cite{Bacchetta:2000jk,thesis}. For sake of 
completeness, we show it in App.~\ref{a:sp} together with the formulae for the pure $s$ 
and $s$-$p$ interference sectors.

\section{Conclusions}
\label{sec:end}

In this paper, we have reconsidered the option of extracting the transversity 
distribution $h_1$ at leading twist by using the analyzing power of the 
interference fragmentation functions (IFF) into two leading unpolarized hadrons inside 
the same current jet. As already shown in Ref.~\cite{Radici:2001na}, in the process 
$ep^{\uparrow} \rightarrow e' h_1 h_2 {X}$ the transversity distribution enters 
a single-spin asymmetry in the azimuthal angle $\phi_R$ of the hadron pair plane. 
The effect survives after the integration upon the transverse component of $P_h=P_1+P_2$. 
Therefore, no transverse-momentum dependent function is required and 
the advantage with respect to the Collins effect is evident. A similar situation was 
known to occurr in the case of fragmentation into spin-1 
hadrons~\cite{Efremov:1982sh,Ji:1994vw,Anselmino:1996vq,Bacchetta:2000jk}, 
but it was never fully examined to the extent of defining a specific asymmetry. 

Here, we have reanalyzed the whole problem in the helicity formalism by further
expanding the IFF in the basis of the relative orbital angular momentum in the cm frame of
the hadron pair. New positivity bounds have been derived. If the invariant mass of the 
pair is not large, the expansion can be limited to the first two modes, namely the 
relative $s$ and $p$ waves. 

Off-diagonal elements in the chirality and in the orbital angular momentum $L$ 
represent the IFF of Ref.~\cite{Jaffe:1998hf} and~\cite{Radici:2001na}, where the 
interference arises from the hadron pair being in a state with either $s$ or $p$ 
relative wave. Elements in the $L=L'=1$ sector correspond to the analysis of 
spin-1 hadron fragmentation~\cite{Bacchetta:2000jk}. Therefore, 
the present formalism represents a unifying framework for the problem of fragmentation 
into two unpolarized hadrons and can be used to correctly and exhaustively
discuss the extraction of transversity from two-hadron leptoproduction.

In fact, after calculating the complete leading-twist cross section, we have identified a 
single spin asymmetry containing two distinct chiral-odd partners of the 
transversity. By integrating the asymmetry over different ranges of the cm polar angle of 
the hadron pair, the transversity $h_1$ can be extracted through the chiral-odd, T-odd 
fragmentation $H_{1,OT}^{\open}$ (corresponding to the $s$-$p$ interference of 
Ref.~\cite{Jaffe:1998hf}) or through the chiral-odd, T-odd fragmentation 
$H_{1,LT}^{\open}$ (corresponding to the $p$-$p$ interference). This second option 
has been often neglected in the literature, despite the fact that the two functions 
have, in principle, a different dependence on the invariant mass and a different
physical origin.

In conclusion, we believe that the fragmentation into two leading unpolarized hadrons can
be a promising tool to measure the transversity distribution, as well as to
achieve further comprehension of the hadronization mechanism.


\begin{acknowledgments}
Several discussions with Daniel Boer are gratefully acknowledged. This work has been 
supported by the TMR network HPRN-CT-2000-00130. 
\end{acknowledgments}


\appendix


\section{}
\label{a:cswithT}

In this appendix, we write explicitly the cross section for two-hadron
leptoproduction at leading order in $1/Q$ and with the inclusion of partonic
transverse momenta. Moreover, we include also T-odd distribution functions,
since recently there have been some indications that they are not forbidden by
time-invariance~\cite{Brodsky:2002cx,Collins:2002kn,Belitsky:2002sm}. 
To simplify the notation, we introduce the projection 
$\vec{a}_T \wedge \vec{b}_T = a_i \eps_T^{ij} b_j$. Inserting in 
Eq.~(\ref{eq:crossdiff}) the formulae for the target helicity density matrix,
Eq.~(\ref{eq:rho}), for the distribution correlation matrix, Eq.~(\ref{eq:phihctw-T}), 
for the elementary scattering matrix, Eq.~(\ref{eq:eqcross}), and the two-hadron 
fragmentation matrix, Eq.~(\ref{eq:deltactw-T}), we obtain the following result:
\begin{eqnarray} 
d^9\!\sigma_{OO}  
&=&
\sum_{a} \frac{\alpha^2 e_a^2}{2\pi Q^2 y}\,\Biggl\{ 
A(y)\,{\mathcal I}\left[f_1 \, D_1\right]
-B(y)\,\frac{|{\vec{R}}_T|}{M_h}\,\cos(\phi_h+\phi_R)\,
   {\mathcal I}\left[\frac{\vec{p}_T \cdot \h}{M}\,
     h_1^{\perp} \, {\bar H}_1^{\open}\right]
\nn \\ && \mbox{} 
  +B(y)\,\frac{|{\vec{R}}_T|}{M_h}\,\sin(\phi_h+\phi_R)\,
   {\mathcal I}\left[\frac{\h \wedge \vec{p}_T}{M}\,
     h_1^{\perp} \, {\bar H}_1^{\open}\right]
\nn \\&&\mbox{}
-B(y)\,\cos(2\phi_h)\,
   {\mathcal I}\left[\frac{2(\vec{p}_T \cdot \h)(\vec{k}_T
 \cdot \h) - \vec{p}_T\cdot \vec{k}_T}{M M_h}\,h_1^{\perp} \, H_1^{\perp}\right]
\nn \\ && \mbox{}
+B(y)\,\sin(2\phi_h)\,
   {\mathcal I}\left[\frac{(\vec{p}_T \cdot \h)(\h \wedge \vec{k}_T) +(\vec{k}_T \cdot 
   \h)(\h \wedge \vec{p}_T)}{M M_h}\,h_1^{\perp} \, H_1^{\perp}\right]\Biggr\},
\label{e:xsectOO} \\ 
d^9\!\sigma_{LO}  
&=& 
- \sum_{a} \frac{\alpha^2 e_a^2}{2\pi Q^2 y}\,\lf\lvert \lambda_e \rg\rvert\,
C(y)\,\frac{|{\vec{R}}_T|}{M_h}\,\Biggl\{\sin(\phi_h-\phi_R)\,
   {\mathcal I}\left[\frac{\vec{k}_T\cdot\h}{M_h}\,
     f_1 \, G_1^{\perp}\right]
\nn \\ && \mbox{} 
+ \cos(\phi_h-\phi_R)\,
   {\mathcal I}\left[\frac{\h \wedge \vec{k}_T}{M_h}\,
     f_1 \, G_1^{\perp}\right]\Biggr\},
\label{e:xsectLO} \\
d^9\!\sigma_{OL}  
&=&
\sum_{a} \frac{\alpha^2 e_a^2}{2\pi Q^2 y}\,\lf\lvert S_L \rg\rvert\Biggl\{
-A(y)\,\frac{|{\vec{R}}_T|}{M_h}\,\sin(\phi_h-\phi_R)\,
   {\mathcal I}\left[\frac{\vec{k}_T\cdot\h}{M_h}\,
     g_{1L} \, G_1^{\perp}\right] 
\nn \\ &&\mbox{}
-A(y)\,\frac{|{\vec{R}}_T|}{M_h}\,\cos(\phi_h-\phi_R)\,
   {\mathcal I}\left[\frac{\h \wedge \vec{k}_T}{M_h}\,
     g_{1L} \, G_1^{\perp}\right] 
\nn \\ &&\mbox{}
+B(y)\,\frac{|{\vec{R}}_T|}{M_h}\,\sin(\phi_h+\phi_R)\,
   {\mathcal I}\left[\frac{\vec{p}_T\cdot\h}{M}\,
     h_{1L}^{\perp} \, {\bar H}_1^{\open}\right]\nn \\
&&\mbox{}
+B(y)\,\frac{|{\vec{R}}_T|}{M_h}\,\cos(\phi_h+\phi_R)\,
   {\mathcal I}\left[\frac{\h \wedge \vec{p}_T}{M}\,
     h_{1L}^{\perp} \, {\bar H}_1^{\open}\right]\nn \\
&&\mbox{}
+B(y)\,\sin(2\phi_h)\,
   {\mathcal I}\left[\frac{2(\vec{p}_T \cdot \h)(\vec{k}_T \cdot \h) 
	- \vec{p}_T\cdot \vec{k}_T}{M M_h}\,
     h_{1L}^{\perp} \, H_1^{\perp}\right]
\nn \\
&&\mbox{}
+B(y)\,\cos(2\phi_h)\,
   {\mathcal I}\left[\frac{(\vec{p}_T \cdot \h)(\h \wedge \vec{k}_T) +(\vec{k}_T \cdot 
   \h)(\h \wedge \vec{p}_T)}{M M_h}\,  h_{1L}^{\perp} \, H_1^{\perp}\right]\Biggr\},
\label{e:xsectOL}
 \\
d^9\!\sigma_{LL}  
&=&
\sum_{a} \frac{\alpha^2 e_a^2}{2\pi Q^2 y}\,\lf\lvert \lambda_e \rg\rvert\,
\lf\lvert S_L \rg\rvert\, C(y)\,
   {\mathcal I}\left[g_{1L} \, D_1\right],
\label{e:xsectLL} 
\end{eqnarray}
\begin{eqnarray}
d^9\!\sigma_{OT}  
&=&
\sum_{a} \frac{\alpha^2 e_a^2}{2\pi Q^2 y}\,\sst\,
A(y)\,\Biggl\{\frac{|{\vec{R}}_T|}{M_h}\,
   \sin(\phi_R-\phi_{S})\,
   {\mathcal I}\left[\frac{\vec{p}_T\cdot \vec{k}_T}{2 M M_h}
     g_{1T} \, G_1^{\perp}\right]
\nn \\ &&\mbox{}
-\frac{|{\vec{R}}_T|}{M_h}\,
   \cos(\phi_R-\phi_{S})\,
   {\mathcal I}\left[\frac{(\vec{p}_T \cdot \h)(\h \wedge \vec{k}_T) -(\vec{k}_T \cdot 
   \h)(\h \wedge \vec{p}_T)}{2 M M_h}  g_{1T} \, G_1^{\perp}\right]
\nn \\ &&\mbox{}
-\frac{|{\vec{R}}_T|}{M_h}\,
   \sin(2\phi_h-\phi_R-\phi_{S})\,
   {\mathcal I}\left[\frac{2(\vec{p}_T \cdot\h)(\vec{k}_T \cdot \h) 
	- \vec{p}_T\cdot \vec{k}_T}{2M M_h}\,
     g_{1T} \, G_1^{\perp}\right]
\nn \\ &&\mbox{}
-\frac{|{\vec{R}}_T|}{M_h}\,
   \cos(2\phi_h-\phi_R-\phi_{S})\,
   {\mathcal I}\left[\frac{(\vec{p}_T \cdot \h)(\h \wedge \vec{k}_T) +(\vec{k}_T \cdot 
   \h)(\h \wedge \vec{p}_T)}{2M M_h}\,  g_{1T} \, G_1^{\perp}\right]\nn \\
&&\mbox{}
+ \sin(\phi_h-\phi_{S})\,
   {\mathcal I}\left[\frac{\vec{p}_T\cdot\h}{M}\,
     f_{1T}^{\perp} \, D_1\right]
+ \cos(\phi_h-\phi_{S})\,
   {\mathcal I}\left[\frac{\h \wedge \vec{p}_T}{M}\,
     f_{1T}^{\perp} \, D_1\right]\Biggr\}
\nn \\ &&\mbox{}
+\sum_{a} \frac{\alpha^2 e_a^2}{2\pi Q^2 y}\,\sst\,
B(y)\,\Biggl\{\sin(\phi_h+\phi_{S})\,
   {\mathcal I}\left[\frac{\vec{k}_T\cdot\h}{M_h}\,
     h_1 \, H_1^{\perp}\right]
\nn \\ &&\mbox{}
+\cos(\phi_h+\phi_{S})\,
   {\mathcal I}\left[\frac{\h \wedge \vec{k}_T}{M_h}\,
     h_1 \, H_1^{\perp}\right]
+\frac{|{\vec{R}}_T|}{M_h}\,\sin(\phi_R+\phi_{S})\,
   {\mathcal I}\left[h_1 \, {\bar H}_1^{\open}\right]
+\sin(3\phi_h-\phi_{S}) \nn \\
&&\mbox{} \times
   {\mathcal I}\left[\frac{4(\vec{p}_T\cdot\h)^2(\vec{k}_T\cdot\h)
	-2(\vec{p}_T\cdot\h)(\vec{p}_T\cdot\vec{k}_T)
	-\vec{p}_T^2(\vec{k}_T\cdot\h)}{2 M^2 M_h}\,
     h_{1T}^{\perp} \, H_1^{\perp}\right]\nn \\
&&\mbox{}
+\cos(3\phi_h-\phi_{S})\, 
   {\mathcal I}\Biggl[\Biggl( \frac{
2(\vec{p}_T\cdot\h)^2(\h \wedge \vec{k}_T)
	+2(\vec{k}_T\cdot\h)(\vec{p}_T\cdot\h)(\h \wedge \vec{p}_T)}{2 M^2 M_h}
\nn \\ && \mbox{} -\frac{
	\vec{p}_T^2(\h \wedge \vec{k}_T)
}{2 M^2 M_h}\Biggr)\,
     h_{1T}^{\perp} \, H_1^{\perp}\Biggr]
+\frac{|{\vec{R}}_T|}{M_h}\,
    \sin(2\phi_h+\phi_R-\phi_{S})\,
   {\mathcal I}\left[\frac{2(\vec{p}_T \cdot \h)^2
		- \vec{p}_T^2}{2 M^2}\,
     h_{1T}^{\perp} \, {\bar H}_1^{\open}\right]
\nn \\
&&\mbox{}
+\frac{|{\vec{R}}_T|}{M_h}\,
    \cos(2\phi_h+\phi_R-\phi_{S})\,
   {\mathcal I}\left[\frac{(\vec{p}_T \cdot \h)(\h \wedge \vec{p}_T)}{2 M^2}\,
     h_{1T}^{\perp} \, {\bar H}_1^{\open}\right]
\Biggr\},
\label{e:xsectOT} \\
d^9\!\sigma_{LT}  
&=&
\sum_{a} \frac{\alpha^2 e_a^2}{2\pi Q^2 y}\,\sst\,C(y)\,\Biggl\{ 
\cos(\phi_h-\phi_{S})\,
   {\mathcal I}\left[\frac{\vec{p}_T\cdot\h}{M}\,
     g_{1T} \, D_1 \right]\nn \\
&&\mbox{}
-\sin(\phi_h-\phi_{S})\,
   {\mathcal I}\left[\frac{\h \wedge \vec{p}_T}{M}\,
     g_{1T} \, D_1 \right]\nn \\
&&\mbox{}
-\frac{|{\vec{R}}_T|}{M_h}\,
    \cos(\phi_R-\phi_{S})\,
   {\mathcal I}\left[\frac{\vec{p}_T\cdot \vec{k}_T}{2 M M_h}
     f_{1T}^{\perp} \, G_1^{\perp}\right]\nn \\
&&\mbox{}
+\frac{|{\vec{R}}_T|}{M_h}\,
    \cos(2\phi_h-\phi_R-\phi_{S})\,
   {\mathcal I}\left[\frac{2(\vec{p}_T \cdot \h)(\vec{k}_T \cdot \h) 
		- \vec{p}_T\cdot \vec{k}_T}{2 M M_h}\,
     f_{1T}^{\perp} \, G_1^{\perp}\right]
\nn \\
&&\mbox{}
-\frac{|{\vec{R}}_T|}{M_h}\,
    \sin(\phi_R-\phi_{S})\,
   {\mathcal I}\left[\frac{(\vec{p}_T \cdot \h)(\h \wedge \vec{k}_T) -(\vec{k}_T \cdot 
   \h)(\h \wedge \vec{p}_T)}{2 M M_h}  f_{1T}^{\perp} \, G_1^{\perp}\right]\nn \\
&&\mbox{}
+\frac{|{\vec{R}}_T|}{M_h}\,
    \sin(2\phi_h-\phi_R-\phi_{S})\,
   {\mathcal I}\left[\frac{(\vec{p}_T \cdot \h)(\h \wedge \vec{k}_T) +(\vec{k}_T \cdot 
   \h)(\h \wedge \vec{p}_T)}{2 M M_h}\,  f_{1T}^{\perp} \, G_1^{\perp}\right]
\Biggr\}.
\label{e:xsectLT}
\end{eqnarray} 
In the case 
of $d^9\sigma_{OT}$, i.e.\ for an unpolarized beam and a transversely polarized 
target, the full expression of the cross section corresponds to the one in Eq.(10) 
of Ref.~\cite{Radici:2001na}, apart for a different overall factor, due to 
slightly different definitions of the hadron tensor and of the fragmentation 
functions, and the use of $M_h$ instead of $M_1 (M_2)$ in the denominators,
due to a different definition of the expansion~(\ref{eq:deltagen}).


\section{}
\label{a:app2}

The full expression of 
$[{\mathcal P}_- \Delta(z,\vec k_T^2,M_h^2) \gamma^-]_{M'M\,\chi_2'\chi_2^{}}^{L'L}$ 
in Eq.~(\ref{eq:deltalmc2-T}) is 
\begin{equation}
  [{\mathcal P}_- \Delta(z,\vec k_T^2,M_h^2) \gamma^-]_{M'M\,\chi_2'\chi_2^{}}^{L'L} =
   \frac{1}{8} \; \left( \begin{array}{cc}
     A_{M'M}^{L'L} & B_{M'M}^{L'L} \\[2pt]
     (B_{M'M}^{L'L})^{\dagger} & C_{M'M}^{L'L} \end{array} \right)  \; ,
\label{eq:deltalmc3-T}
\end{equation}
where
\bea
   A_{M'M}^{L'L} &= &{} \nn \\[4pt]
   & &\hspace{-2truecm} \left( \begin{array}{c|ccc}
       \scriptstyle{D_{1,OO}^s} & \scriptstyle{-\sqrt{\frac{2}{3}}\,e^{i\phi}\,
       (D_{1,OT}+i\frac{|\vec k_T||\vec R|}{M_h^2}\, G_{1,OT}^\perp)} & 
       \scriptstyle{\frac{2}{\sqrt{3}} \, D_{1,OL}} & \scriptstyle{\sqrt{\frac{2}{3}}
       \,e^{-i\phi}\,(D_{1,OT}-i\frac{|\vec k_T||\vec R|}{M_h^2}\, G_{1,OT}^\perp)} 
       \\[7pt]
       \hline \\[-12pt]
       \scriptstyle{-\sqrt{\frac{2}{3}}\,e^{-i\phi}\,(D_{1,OT}-
       i\frac{|\vec k_T||\vec R|}{M_h^2}\, G_{1,OT}^\perp)} & 
       \scriptstyle{D_{1,OO}^p-\frac{1}{3}\,D_{1,LL}} & 
       \scriptstyle{-\frac{\sqrt{2}}{3}\,e^{-i\phi}\,(D_{1,LT}-
       i\frac{|\vec k_T||\vec R|}{M_h^2}\, G_{1,LT}^\perp)} & 
       \scriptstyle{-\frac{2}{3}\,e^{-2i\phi}\,(2D_{1,TT}- 
       i\frac{|\vec k_T||\vec R|}{M_h^2}\, G_{1,TT}^\perp) }\\[2pt]
       \scriptstyle{\frac{2}{\sqrt{3}}\,D_{1,OL}} & 
       \scriptstyle{-\frac{\sqrt{2}}{3}\,e^{i\phi}\,(D_{1,LT}+
       i\frac{|\vec k_T||\vec R|}{M_h^2}\, G_{1,LT}^\perp)} & 
       \scriptstyle{D_{1,OO}^p+\frac{2}{3}\,D_{1,LL}} & 
       \scriptstyle{\frac{\sqrt{2}}{3}\,e^{-i\phi}\,(D_{1,LT}-
       i\frac{|\vec k_T||\vec R|}{M_h^2}\, G_{1,LT}^\perp)} \\[2pt]
       \scriptstyle{\sqrt{\frac{2}{3}}\,e^{i\phi}\,(D_{1,OT}+
       i\frac{|\vec k_T||\vec R|}{M_h^2}\, G_{1,OT}^\perp)} & 
       \scriptstyle{-\frac{2}{3}\,e^{2i\phi}\,(2D_{1,TT}+ 
       i\frac{|\vec k_T||\vec R|}{M_h^2}\, G_{1,TT}^\perp)} & 
       \scriptstyle{\frac{\sqrt{2}}{3}\,e^{i\phi}\,
       (D_{1,LT}+i\frac{|\vec k_T||\vec R|}{M_h^2}\, G_{1,LT}^\perp)} & 
       \scriptstyle{D_{1,OO}^p - \frac{1}{3}\,D_{1,LL}} \end{array} \right)  
       \nn \\[2pt]
& &{} \label{eq:almc-T} \\[10pt]
   B_{M'M}^{L'L} &= &{} \nn \\[4pt] 
   & &\hspace{-1truecm} i\frac{|\vec k_T|}{M_h}\; \left( 
     \begin{array}{c|ccc} 
      \scriptstyle{e^{i\phi}\,H_{1,OO}^{\perp\,s}} & 
      \scriptstyle{-\frac{2\sqrt{2}}{\sqrt{3}}\,e^{2i\phi}\,
      H_{1,OT}^\perp} & \scriptstyle{\frac{2}{\sqrt{3}}\,e^{i\phi}\,H_{1,OL}^\perp} & 
      \scriptstyle{\frac{2\sqrt{2}}{\sqrt{3}} \,(\frac{|\vec R|}{|\vec k_T|}\, 
      {\bar H}_{1,OT}^{\open}+ H_{1,OT}^\perp)} \\[7pt]
      \hline \\[-12pt]
      \scriptstyle{-\frac{2\sqrt{2}}{\sqrt{3}} \,(\frac{|\vec R|}{|\vec k_T|}\, 
      {\bar H}_{1,OT}^{\open}+ H_{1,OT}^\perp)} & 
      \scriptstyle{e^{i\phi}\,(H_{1,OO}^{\perp\,p}-\frac{1}{3}\,H_{1,LL}^\perp)} & 
      \scriptstyle{-\frac{2\sqrt{2}}{3} \,(\frac{|\vec R|}{|\vec k_T|}\, 
      {\bar H}_{1,LT}^{\open}+H_{1,LT}^\perp)} & 
      \scriptstyle{-\frac{8}{3}\,e^{i\phi}\,(\frac{|\vec R|}{|\vec k_T|}\, 
      H_{1,TT}^{\open}+H_{1,TT}^\perp)} \\[2pt]
      \scriptstyle{\frac{2}{\sqrt{3}}\,e^{i\phi}\,H_{1,OL}^\perp} &
      \scriptstyle{-\frac{2\sqrt{2}}{\sqrt{3}}\,e^{2i\phi}\,H_{1,TT}^\perp} & 
      \scriptstyle{e^{i\phi}\,(H_{1,OO}^{\perp\,p}+\frac{2}{3}\,H_{1,LL}^\perp)} & 
      \scriptstyle{\frac{2\sqrt{2}}{3} \,(\frac{|\vec R|}{|\vec k_T|}\, 
      {\bar H}_{1,LT}^{\open} + H_{1,LT}^\perp)} \\[2pt]
      \scriptstyle{\frac{2\sqrt{2}}{\sqrt{3}}\,e^{2i\phi}\,H_{1,OT}^\perp} &
      \scriptstyle{-\frac{8}{3}\,e^{3i\phi}\,H_{1,TT}^\perp} & 
      \scriptstyle{\frac{2\sqrt{2}}{\sqrt{3}}\,e^{2i\phi}\, H_{1,TT}^\perp} & 
      \scriptstyle{e^{i\phi}\,(H_{1,OO}^{\perp\,p}-\frac{1}{3}\,
      H_{1,LL}^\perp)} \end{array} \right) \; , \nn \\
& &{} \label{eq:blmc-T}
\eea
and $\phi\equiv \phi_k-\phi_R$. The matrix~(\ref{eq:deltalmc3-T}) respects 
Hermiticity, angular momentum conservation, and parity 
invariance. Due to the explicit dependence upon the transverse momentum $\vec k_T$, the 
conditions for angular momentum and parity conservation read
\bea 
M+\chi'_{2} &= &M'+\chi_{2}+l_{k_T} \nn \\ 
\left[ {\cal P}_- \Delta \g^- \right]_{M'M\,\chi_2'\chi_2^{}}^{L'L} &= &(-1)^{l_{k_T}}
\left[ {\cal P}_- \Delta \g^- \right]_{-M'-M\,-\chi_2'-\chi_2^{}}^{L'L} \; ,
\eea
where $l_{k_T}$ denotes the units of angular momentum introduced by $\vec k_T$. 
From the last constraint it is possible to derive the lower right block, 
i.e.\ $C_{M'M}^{L'L} = (-1)^{l_{k_T}} \,A_{-M'-M}^{L'L}$. 

Again, as in the case of Eq.~(\ref{eq:dOO}), we have 
\begin{equation}
  H_{1,OO}(z,\vec k_T^2,M_h^2) = \frac{1}{4}\, 
  H_{1,OO}^{\perp\,s}(z,\vec k_T^2,M_h^2) + 
  \frac{3}{4}\,H_{1,OO}^{\perp\,p}(z,\vec k_T^2,M_h^2) 
\label{eq:hOO}
\end{equation}
and the functions $H_{1,OO}^{\perp\,s}, H_{1,OO}^{\perp\,p}$ are kinematically 
indistinguishable unless some hypothesis is made on their $M_h^2$ dependence. The
$L=L'=1$ sector of Eqs.~(\ref{eq:almc-T},\ref{eq:blmc-T}) has been studied in the
case of spin-1 fragmentation~\cite{Bacchetta:2000jk}. The interference $(L=0,L'=1)$ 
sector has never been analyzed in this form, namely including the explicit dependence 
on $\vec k_T$. Finally, from 
$[{\mathcal P}_- \Delta \gamma^-]_{M'M\,\chi_2'\chi_2^{}}^{L'L}$ 
being positive semidefinite, it is possible to derive bounds on each of the displayed 
fragmentation functions.

 
\section{}
\label{a:sp}
In this appendix, we explicitly present the complete cross section for the production
of two unpolarized hadrons in relative $s$ and $p$ waves, at leading order in 
$1/Q$, including transverse momenta and T-odd distribution and fragmentation
functions. 

The cross section is obtained by replacing
Eqs.~(\ref{eq:rho},\ref{eq:phihctw-T},\ref{eq:eqcross},\ref{eq:deltalmc3-T},
\ref{eq:decay}) in Eq.~(\ref{eq:crossdiffhpw}). It is convenient to introduce the 
following combination of fragmentation functions
\begin{align} 
H_{1,OT}^{\open}&=
\bar{H}_{1,OT}^{\open}+\frac{|\vec{k}_T|}{\rr} H_{1,OT}^{\perp}, \\
H_{1,LT}^{\open}&=
\bar{H}_{1,LT}^{\open}+\frac{|\vec{k}_T|}{\rr} H_{1,LT}^{\perp}, \\
H_{1,TT}^{\open}&=
\bar{H}_{1,TT}^{\open}+\frac{|\vec{k}_T|}{\rr} H_{1,TT}^{\perp}. 
\end{align}


\subsubsection{Unpolarized lepton beam and unpolarized target} 

\begin{equation} 
\begin{split}
\label{e:crosssecOO}
d^8\!\sigma_{OO}  
&=
\sum_{a} \frac{\alpha^2 e_a^2}{2 \pi s x y^2}\,
A(y)\,\Biggl\{{\cal I}\left[f_1 \, \left(\frac{1}{4} D_{1,OO}^s 
				+ \frac{3}{4}D_{1,OO}^p\right)\right]
+ \cos \theta\, {\cal I}\left[f_1 \, D_{1,OL} \right]
\\ 
& \quad
+ \frac{1}{3}\lf(3 \cos^2{\theta} -1\rg) {\cal I}\left[f_1 \, \left(\frac{3}{4} D_{1,LL} 
\right)\right] + \sin \theta\, \cos(\phi_h - \phi_R) \,{\cal I}\left[
\frac{\vec{k}_T\cdot\h}{M_h}\,f_1 \, \lf(-\frac{M_h}{|\vec{k}_T|}D_{1,OT}\rg)\right] \\ 
& \quad - \sin 2\theta\, \cos(\phi_h - \phi_R) \,{\cal I}\left[
\frac{\vec{k}_T\cdot\h}{M_h}\,f_1 \, \lf(-\frac{M_h}{2|\vec{k}_T|}D_{1,LT}\rg)\right]
- \sin^2\theta\, \cos(2\phi_h - 2\phi_R) \\ 
& \quad \times {\cal I}\left[\frac{2(\vec{k}_T \cdot \h)^2- \vec{k}_T^2}{M_h^2} \,f_1 \, 
\lf(-\frac{M_h^2}{|\vec{k}_T|^2}D_{1,TT} \rg)\right]\Biggr\} \\
& \quad +\sum_{a} \frac{\alpha^2 e_a^2}{2 \pi s x y^2}\, B(y)\,\Biggl\{-\cos 2\phi_h
{\cal I}\left[
\frac{2(\vec{p}_T \cdot \h)(\vec{k}_T\cdot \h) - \vec{p}_T\cdot \vec{k}_T}{M M_h}\,  
h_1^{\perp} \, \lf(\frac{1}{4} H_{1,OO}^{\perp\,s}+\frac{3}{4} H_{1,OO}^{\perp\,p}
\rg)\right] \\
& \quad
-\frac{1}{3}\lf(3 \cos^2{\theta} -1\rg) \cos 2\phi_h {\cal I} \left[
\frac{2(\vec{p}_T \cdot \h)(\vec{k}_T \cdot \h) - \vec{p}_T\cdot \vec{k}_T}{M M_h}\,  
h_1^{\perp} \, \lf(\frac{3}{4} H_{1,LL}^{\perp} \rg)\right] \\ 
& \quad +\sin \theta\, \cos (\phi_h + \phi_R){\cal I}\left[
\frac{\vec{p}_T\cdot\h}{M}\, h_1^{\perp}\,\lf(-\frac{|\vec{R}|}{M_h}H_{1,OT}^{\open} \rg)
\right] +\sin 2\theta\, \cos (\phi_h + \phi_R) \\ 
& \quad \times {\cal I}\left[\frac{\vec{p}_T\cdot\h}{M}\, h_1^{\perp}\, 
\lf(-\frac{|\vec{R}|}{2 M_h}H_{1,LT}^{\open} \rg)\right]
+\sin^2\theta\, \cos 2\phi_R \,{\cal I}\left[
\frac{\vec{p}_T\cdot\vec{k}_T}{M M_h}\, h_1^{\perp}\,
\lf( -\frac{|\vec{R}|}{|\vec{k}_T|}H_{1,TT}^{\open}\rg)\right] \\ 
& \quad +\sin \theta\, \cos (3 \phi_h - \phi_R) \\ 
& \quad \times {\cal I}\left[\frac{4(\vec{k}_T\cdot\h)^2(\vec{p}_T\cdot\h)
	-2(\vec{k}_T\cdot\h)(\vec{p}_T\cdot\vec{k}_T)
	-\vec{k}_T^2(\vec{p}_T\cdot\h)}{2 M M_h^2}\,
     h_1^{\perp} \, \lf(-\frac{2 M_h}{|\vec{k}_T|}H_{1,OT}^{\perp} \rg)\right] \\ 
& \quad +\sin 2\theta\, \cos (3 \phi_h - \phi_R) \\ 
& \quad \times 
	{\cal I}\left[\frac{4(\vec{k}_T\cdot\h)^2(\vec{p}_T\cdot\h)
	-2(\vec{k}_T\cdot\h)(\vec{p}_T\cdot\vec{k}_T)
	-\vec{k}_T^2(\vec{p}_T\cdot\h)}{2 M M_h^2}\,
     h_1^{\perp} \, \lf(-\frac{M_h}{|\vec{k}_T|}H_{1,LT}^{\perp} \rg)\right] \\ 
& \quad +\sin^2 \theta\, \cos (4 \phi_h - 2 \phi_R) \,{\cal I}\Biggl[
\Biggl( \frac{\lf[\vec{k}_T^2 -4(\vec{k}_T\cdot\h)^2\rg]\lf[\vec{p}_T\cdot\vec{k}_T-
4(\vec{k}_T\cdot\h)(\vec{p}_T\cdot\h)\rg]}{2 M M_h^3} \\ 
& \quad
	-\frac{8(\vec{k}_T\cdot\h)^3(\vec{p}_T\cdot\h)}{2 M M_h^3}\Biggr)\,
     h_1^{\perp} \, \lf(-\frac{2 M_h^2}{|\vec{k}_T|^2}H_{1,TT}^{\perp} \rg)\Biggr]
\Biggr\} \; ,
\end{split}
\end{equation}


\subsubsection{Polarized lepton beam and unpolarized target} 

\begin{equation} 
\begin{split} 
\label{e:crosssecLO}
d^8\!\sigma_{LO}  
&= 
- \sum_{a} \frac{\alpha^2 e_a^2}{2 \pi s x y^2}\,\lambda_e\,
C(y)\,\Biggl\{
\sin \theta\, \sin(\phi_h - \phi_R)\,{\cal I}\left[\frac{\vec{k}_T\cdot\h}{M_h}\,f_1 \, 
\lf(\frac{|\vec{R}|}{ M_h} G_{1,OT}^{\perp} \rg)\right] \\ 
& \quad +\sin 2\theta\, \sin(\phi_h - \phi_R) \,{\cal I}\left[
\frac{\vec{k}_T\cdot\h}{M_h}\,f_1 \, 
\lf(\frac{|\vec{R}|}{2 M_h} G_{1,LT}^{\perp} \rg)\right] \\ 
& \quad +\sin^2 \theta\, \sin(2\phi_h - 2\phi_R) \,{\cal I}\left[
\frac{2(\vec{k}_T \cdot \h)^2- \vec{k}_T^2}{M_h^2}\, f_1 \, 
\lf(\frac{|\vec{R}|}{2|\vec{k}_T|} G_{1,TT}^{\perp} \rg)\right]\Biggr\} \; ,
\end{split}
\end{equation}


\subsubsection{Unpolarized lepton beam and longitudinally polarized target} 

\begin{equation} 
\begin{split}
\label{e:crosssecOL}
d^8\!\sigma_{OL}  
&=
-\sum_{a} \frac{\alpha^2 e_a^2}{2 \pi s x y^2}\,\lf\lvert S_L \rg\rvert
\,A(y)\,
\Biggl\{\sin \theta \,\sin(\phi_h-\phi_R)\,
   {\cal I}\left[\frac{\vec{k}_T\cdot\h}{M_h}\,
     g_{1L} \, \lf(\frac{|\vec{R}|}{M_h} G_{1,OT}^{\perp} \rg)\right]
\\ 
& \quad 
+\sin 2\theta \,\sin(\phi_h-\phi_R)\,
   {\cal I}\left[\frac{\vec{k}_T\cdot\h}{M_h}\,
     g_{1L} \, \lf(\frac{|\vec{R}|}{2 M_h} G_{1,LT}^{\perp} \rg)\right]
\\ 
& \quad 
+ \sin^2 \theta\, \sin(2\phi_h - 2\phi_R)
	\,{\cal I}\left[\frac{2(\vec{k}_T \cdot \h)^2
		- \vec{k}_T^2}{M_h^2}\,
     g_{1L} \, \lf(\frac{|\vec{R}|}{2|\vec{k}_T|} G_{1,TT}^{\perp} \rg)\right] 
\Biggr\} \\ 
& \quad
-\sum_{a} \frac{\alpha^2 e_a^2}{2 \pi s x y^2}\,\lf\lvert S_L \rg\rvert
\,B(y)\,\Biggl\{\sin 2\phi_h{\cal I} \left[
\frac{2(\vec{p}_T \cdot \h)(\vec{k}_T \cdot \h) - \vec{p}_T\cdot \vec{k}_T}{M M_h}\,  
h_{1L}^{\perp} \,
\lf(\frac{1}{4} H_{1,OO}^{\perp\,s}+\frac{3}{4} H_{1,OO}^{\perp\,p} \rg)\right] \\ 
& \quad -\frac{1}{3}\lf(3 \cos^2{\theta} -1\rg)\, \sin 2\phi_h \times {\cal I} \left[
\frac{2(\vec{p}_T \cdot \h)(\vec{k}_T \cdot \h) - \vec{p}_T\cdot \vec{k}_T}{M M_h}\,  
h_{1L}^{\perp} \, \lf(\frac{3}{4} H_{1,LL}^{\perp} \rg)\right] \\ 
& \quad +\sin \theta \,\sin(\phi_h+\phi_R) {\cal I}\left[\frac{\vec{p}_T\cdot\h}{M}\,
h_{1L}^{\perp} \, \lf(-\frac{|\vec{R}|}{M_h}H_{1,OT}^{\open} \rg)\right] 
+\sin 2\theta \,\sin(\phi_h+\phi_R) \\ 
& \quad \times {\cal I}\left[\frac{\vec{p}_T\cdot\h}{M}\,
h_{1L}^{\perp} \, \lf(-\frac{|\vec{R}|}{2 M_h}H_{1,LT}^{\open} \rg)\right] 
+ \sin^2 \theta\, \sin 2\phi_R \,{\cal I}\left[\frac{\vec{p}_T\cdot\vec{k}_T}{M M_h}\,
h_{1L}^{\perp} \, \lf( -\frac{|\vec{R}|}{|\vec{k}_T|}H_{1,TT}^{\open}\rg)\right] \\ 
& \quad  +\sin \theta \,\sin(3 \phi_h - \phi_R) \\ 
& \quad \times   {\cal I}\left[
\frac{4(\vec{k}_T\cdot\h)^2(\vec{p}_T\cdot\h)-2(\vec{k}_T\cdot\h)(\vec{p}_T\cdot\vec{k}_T)
      -\vec{k}_T^2(\vec{p}_T\cdot\h)}{2 M M_h^2}\,
h_{1L}^{\perp} \, \lf(-\frac{2 M_h}{|\vec{k}_T|}H_{1,OT}^{\perp} \rg)\right] \\ 
& \quad  +\sin 2\theta \,\sin(3 \phi_h - \phi_R) \\ 
& \quad \times {\cal I}\left[
\frac{4(\vec{k}_T\cdot\h)^2(\vec{p}_T\cdot\h)-2(\vec{k}_T\cdot\h)(\vec{p}_T\cdot\vec{k}_T)
      -\vec{k}_T^2(\vec{p}_T\cdot\h)}{2 M M_h^2}\,
h_{1L}^{\perp} \, \lf(-\frac{M_h}{|\vec{k}_T|}H_{1,LT}^{\perp} \rg)\right] \\ 
& \quad +\sin^2 \theta\, \sin (4 \phi_h - 2 \phi_R) {\cal I}\Biggl[ \Biggl(
\frac{\lf[\vec{k}_T^2 -4(\vec{k}_T\cdot\h)^2\rg]\lf[\vec{p}_T\cdot\vec{k}_T
       -4(\vec{k}_T\cdot\h)(\vec{p}_T\cdot\h)\rg]}{2 M M_h^3} \\ 
& \quad -\frac{8(\vec{k}_T\cdot\h)^3(\vec{p}_T\cdot\h)}{2 M M_h^3}\Biggr)\,
h_{1L}^{\perp} \, \lf(-\frac{2 M_h^2}{|\vec{k}_T|^2}H_{1,TT}^{\perp} \rg)\Biggr]
\Biggr\} \; ,
\end{split}
\end{equation}


\subsubsection{Polarized lepton beam and longitudinally polarized target} 

\begin{equation} 
\begin{split}
\label{e:crosssecLL}
d^8\!\sigma_{LL}  
&=
\sum_{a} \frac{\alpha^2 e_a^2}{2 \pi s x y^2}\,\lambda_e\, \lf\lvert S_L \rg\rvert\, 
C(y)\, \Biggl\{{\cal I}\left[g_{1L} \, \left(\frac{1}{4} D_{1,OO}^s + \frac{3}{4} 
D_{1,OO}^p\right)\right]+ \cos \theta\, {\cal I}\left[g_{1L} \, D_{1,OL} \right] \\
& \quad + \frac{1}{3}\lf(3 \cos^2{\theta} -1\rg) {\cal I}\left[g_{1L} \, \left(
\frac{3}{4} D_{1,LL} \right)\right] \\ 
& \quad - \sin \theta\, \cos(\phi_h - \phi_R) \,{\cal I}\left[
\frac{\vec{k}_T\cdot\h}{M_h}\,g_{1L} \, \lf(-\frac{M_h}{|\vec{k}_T|}D_{1,LT}\rg)\right]
\\ 
& \quad - \sin 2\theta\, \cos(\phi_h - \phi_R) \,{\cal I}\left[
\frac{\vec{k}_T\cdot\h}{M_h}\,g_{1L} \, \lf(-\frac{M_h}{2|\vec{k}_T|}D_{1,LT}\rg)\right]
\\ 
& \quad - \sin^2\theta\, \cos(2\phi_h - 2\phi_R) \,{\cal I}\left[
\frac{2(\vec{k}_T \cdot \h)^2- \vec{k}_T^2}{M_h^2} \,g_{1L} \, 
\lf(-\frac{M_h^2}{|\vec{k}_T|^2}D_{1,TT} \rg)\right]\Biggr\} \; ,
\end{split} 
\end{equation}


\subsubsection{Unpolarized lepton beam and  transversely polarized target} 

\begin{align} 
\label{e:crosssecOT}
d^8\!\sigma_{OT}  
&=
\sum_{a} \frac{\alpha^2 e_a^2}{2 \pi s x y^2}\,\sst\,A(y)\,
\Biggl\{
\sin \theta \sin(\phi_R - \phi_S)
	\,{\cal I}\left[\frac{(\vec{p}_T \cdot \vec{k}_T)}{2 M M_h} 
		\,g_{1T} \, \lf(\frac{|\vec{R}|}{ M_h} G_{1,OT}^{\perp} \rg)\right]
\nn \\ & \quad
-\sin \theta \sin(2 \phi_h - \phi_R - \phi_S)
	\,{\cal I}\left[\frac{2(\vec{p}_T \cdot \h)(\vec{k}_T
 \cdot \h) - \vec{p}_T\cdot \vec{k}_T}{2 M M_h}
		\,g_{1T} \, \lf(\frac{|\vec{R}|}{ M_h} G_{1,OT}^{\perp} \rg)\right]
\nn \\ & \quad
+\sin 2\theta \sin(\phi_R - \phi_S)
	\,{\cal I}\left[\frac{(\vec{p}_T \cdot \vec{k}_T)}{2 M M_h} 
		\,g_{1T} \, \lf(\frac{|\vec{R}|}{2 M_h} G_{1,LT}^{\perp} \rg)\right]
\nn \\ & \quad
-\sin 2\theta \sin(2 \phi_h - \phi_R - \phi_S)
	\,{\cal I}\left[\frac{2(\vec{p}_T \cdot \h)(\vec{k}_T
 \cdot \h) - \vec{p}_T\cdot \vec{k}_T}{2 M M_h}
		\,g_{1T} \, \lf(\frac{|\vec{R}|}{2 M_h} G_{1,LT}^{\perp} \rg)\right]
\nn \\ & \quad
-\sin^2 \theta \sin(\phi_h - 2 \phi_R + \phi_S)
	\,{\cal I}\left[\frac{2 (\vec{k}_T\cdot\h) (\vec{p}_T\cdot\vec{k}_T)
	- \vec{k}_T^2(\vec{p}_T\cdot\h)}{2 M M_h^2} 
		\,g_{1T} \, \lf(\frac{|\vec{R}|}{2|\vec{k}_T|} G_{1,TT}^{\perp} \rg)
		\right]
\nn \\ & \quad
-\sin^2 \theta \sin(3 \phi_h - 2 \phi_R - \phi_S)
	\,{\cal I}\Biggl[\Biggl(\frac{4(\vec{k}_T\cdot\h)^2(\vec{p}_T\cdot\h)
	-2(\vec{k}_T\cdot\h)(\vec{p}_T\cdot\vec{k}_T)}{2 M M_h^2}
\nn \\ & \quad 
	-\frac{\vec{k}_T^2(\vec{p}_T\cdot\h)}{2 M M_h^2}\Biggr)
		\,g_{1T} \, \lf(\frac{|\vec{R}|}{2|\vec{k}_T|} G_{1,TT}^{\perp} \rg)
		\Biggr]
+\sin(\phi_h - \phi_S)
	\,{\cal I}\left[\frac{\vec{p}_T \cdot \h}{M}
		\,f_{1T}^{\perp} \, \left(\frac{1}{4} D_{1,OO}^s + \frac{3}{4}
D_{1,OO}^p\right)\right]
\nn \\
& \quad
+ \cos \theta\, {\cal I}\left[f_{1T}^{\perp} \, D_{1,OL} \right]
+\frac{1}{3}\lf(3 \cos^2{\theta} -1\rg) 
\sin(\phi_h - \phi_S)
	\,{\cal I}\left[\frac{\vec{p}_T \cdot \h}{M}
		\,f_{1T}^{\perp} \, \left(\frac{3}{4} D_{1,LL} \right)\right]
\nn \\ 
& \quad
-\sin \theta \sin(\phi_R - \phi_S)
	\,{\cal I}\left[\frac{(\vec{p}_T \cdot \vec{k}_T)}{2 M M_h} 
		\,f_{1T}^{\perp} \, \lf(-\frac{M_h}{|\vec{k}_T|}D_{1,OT}\rg)\right]
\nn \\ & \quad
-\sin \theta \sin(2 \phi_h - \phi_R - \phi_S)
	\,{\cal I}\left[\frac{2(\vec{p}_T \cdot \h)(\vec{k}_T
 \cdot \h) - \vec{p}_T\cdot \vec{k}_T}{2 M M_h}
		\,f_{1T}^{\perp} \, \lf(-\frac{M_h}{|\vec{k}_T|}D_{1,OT}\rg)\right]
\nn \\ & \quad
-\sin 2\theta \sin(\phi_R - \phi_S)
	\,{\cal I}\left[\frac{(\vec{p}_T \cdot \vec{k}_T)}{2 M M_h} 
		\,f_{1T}^{\perp} \, \lf(-\frac{M_h}{2|\vec{k}_T|}D_{1,LT}\rg)\right]
\nn \\ & \quad
-\sin 2\theta \sin(2 \phi_h - \phi_R - \phi_S)
	\,{\cal I}\left[\frac{2(\vec{p}_T \cdot \h)(\vec{k}_T
 \cdot \h) - \vec{p}_T\cdot \vec{k}_T}{2 M M_h}
		\,f_{1T}^{\perp} \, \lf(-\frac{M_h}{2|\vec{k}_T|}D_{1,LT}\rg)\right]
\nn \\ & \quad
+\sin^2 \theta \sin(\phi_h - 2 \phi_R + \phi_S)
	\,{\cal I}\left[\frac{2 (\vec{k}_T\cdot\h) (\vec{p}_T\cdot\vec{k}_T)
	- \vec{k}_T^2(\vec{p}_T\cdot\h)}{2 M M_h^2} 
		\,f_{1T}^{\perp} \, \lf(-\frac{M_h^2}{|\vec{k}_T|^2}D_{1,TT} \rg)\right]
\nn \\ & \quad
-\sin^2 \theta \sin(3 \phi_h - 2\phi_R - \phi_S)
	\,{\cal I}\Biggl[\Biggl(\frac{4(\vec{k}_T\cdot\h)^2(\vec{p}_T\cdot\h)
	-2(\vec{k}_T\cdot\h)(\vec{p}_T\cdot\vec{k}_T)}{2 M M_h^2}
\nn \\ & \quad 
	-\frac{\vec{k}_T^2(\vec{p}_T\cdot\h)}{2 M M_h^2}\Biggr)
		\,f_{1T}^{\perp} \, \lf(-\frac{M_h^2}{|\vec{k}_T|^2}D_{1,TT} \rg)\Biggr]
		\Biggr\}
\nn \\ & \quad
+\sum_{a}\frac{\alpha^2 e_a^2}{2 \pi s x y^2}\,
B(y)\,
\Biggl\{\cos 2\phi_h{\cal I}\left[\frac{2(\vec{p}_T \cdot \h)(\vec{k}_T
 \cdot \h) - \vec{p}_T\cdot \vec{k}_T}{M M_h}\,  h_1 \,
\lf(\frac{1}{4} H_{1,OO}^{\perp\,s}+\frac{3}{4} H_{1,OO}^{\perp\,p}
\rg)\right] \nn \\ 
& \quad
+\frac{1}{3}\lf(3 \cos^2{\theta} -1\rg) \sin(\phi_h + \phi_S){\cal I} \left[
\frac{\vec{k}_T \cdot \h}{M_h}\,  h_1 \, \lf(\frac{3}{4} H_{1,LL}^{\perp} \rg)\right] 
\nn \\ & \quad
-\sin \theta\, \sin (\phi_R + \phi_S) 
	\,{\cal I}\left[h_1\, \lf(-\frac{|\vec{R}|}{M_h}H_{1,OT}^{\open}
\rg)\right]
\nn \\
& \quad
-\sin 2\theta\, \sin (\phi_R + \phi_S) 
	\,{\cal I}\left[h_1\, \lf(-\frac{|\vec{R}|}{2 M_h}H_{1,LT}^{\open} \rg)\right]
\\ 
& \quad
+\sin^2\theta\,
\sin (\phi_h -2\phi_R-\phi_S) 
	\,{\cal I}\left[\frac{\vec{k}_T \cdot \h}{M_h}\, h_1\,
\lf( -\frac{|\vec{R}|}{|\vec{k}_T|}H_{1,TT}^{\open}\rg)\right]
\nn \\ & \quad
-\sin \theta\, 
\sin (2 \phi_h - \phi_R+ \phi_S) 
	{\cal I}\left[\frac{2(\vec{k}_T \cdot \h)^2
		- \vec{k}_T^2}{2 M_h^2} \,
     h_1 \, \lf(-\frac{2 M_h}{|\vec{k}_T|}H_{1,OT}^{\perp} \rg)\right]
\nn \\ 
& \quad
-\sin 2\theta\, 
\sin (2 \phi_h - \phi_R+ \phi_S) 
	{\cal I}\left[\frac{2(\vec{k}_T \cdot \h)^2
		- \vec{k}_T^2}{2 M_h^2} \,
     h_1 \, \lf(-\frac{M_h}{|\vec{k}_T|}H_{1,LT}^{\perp} \rg)\right]
\nn \\ & \quad
-\sin^2 \theta\, \sin (3 \phi_h - 2 \phi_R+\phi_S) 
	{\cal I}\left[\frac{4(\vec{k}_T \cdot \h)^3
		- 3 \vec{k}_T^2 (\vec{k}_T \cdot \h)}{2 M_h^3} \,
     h_1 \, \lf(-\frac{2 M_h^2}{|\vec{k}_T|^2}H_{1,TT}^{\perp} \rg)\right]
\nn \\ & \quad
+\cos 2\phi_h{\cal I}\left[\frac{2(\vec{p}_T \cdot \h)(\vec{k}_T
 \cdot \h) - \vec{p}_T\cdot \vec{k}_T}{M M_h}\,  h_{1T}^{\perp} \,
\lf(\frac{1}{4} H_{1,OO}^{\perp\,s}+\frac{3}{4} H_{1,OO}^{\perp\,p}
\rg)\right]
\nn \\ & \quad
+\frac{1}{3}\lf(3 \cos^2{\theta} -1\rg) \sin(3 \phi_h - \phi_S)
\nn \\ & \quad \times 
{\cal I} \left[\frac{4(\vec{p}_T\cdot\h)^2(\vec{k}_T\cdot\h)-2(\vec{p}_T\cdot\h)
(\vec{p}_T\cdot\vec{k}_T)-\vec{p}_T^{\, 2}(\vec{k}_T\cdot\h)}{2 M^2 M_h} \,  
h_{1T}^{\perp} \, \lf(\frac{3}{4} H_{1,LL}^{\perp} \rg)\right] 
\nn \\ & \quad
-\sin \theta\, \sin (2 \phi_h +\phi_R - \phi_S) \,{\cal I}\left[ 
\frac{2(\vec{p}_T \cdot \h)^2- \vec{p}_T^{\, 2}}{2 M^2} \, h_{1T}^{\perp}\, 
\lf(-\frac{|\vec{R}|}{ M_h}H_{1,OT}^{\open} \rg)\right]
\nn \\ & \quad
-\sin 2\theta\, \sin (2 \phi_h +\phi_R - \phi_S) \,{\cal I}\left[ 
\frac{2(\vec{p}_T \cdot \h)^2- \vec{p}_T^{\, 2}}{2 M^2} \, h_{1T}^{\perp}\, 
\lf(-\frac{|\vec{R}|}{2 M_h}H_{1,LT}^{\open} \rg)\right]
\nn \\ & \quad
-\sin^2\theta\, \sin (\phi_h +2\phi_R-\phi_S) \,{\cal I}\left[
\frac{2 (\vec{p}_T \cdot \vec{k}_T)(\vec{p}_T \cdot \h)- (\vec{k}_T \cdot \h)
\vec{p}_T^{\, 2}}{2 M^2 M_h}  
 \, h_{1T}^{\perp}\,\lf( -\frac{|\vec{R}|}{|\vec{k}_T|}H_{1,TT}^{\open}\rg)\right]
\nn \\ & \quad
+\sin \theta\, \sin (4 \phi_h - \phi_R- \phi_S) \,{\cal I}\Biggl[\Biggl(
\frac{\vec{k}_T^2\lf[2(\vec{p}_T \cdot \h)^2- \vec{p}_T^{\, 2}\rg]}{4 M^2 M_h^2} - 
2 (\vec{k}_T \cdot \h) \nn \\ & \quad \times
\frac{\lf[4(\vec{p}_T\cdot\h)^2(\vec{k}_T\cdot\h)-2(\vec{p}_T\cdot\h)
(\vec{p}_T\cdot\vec{k}_T)-\vec{p}_T^{\, 2}(\vec{k}_T\cdot\h)
\rg]}{4 M^2 M_h^2}\Biggr)  \,
 h_{1T}^{\perp} \, \lf(-\frac{2 M_h}{|\vec{k}_T|}H_{1,OT}^{\perp} \rg)\Biggr]
\nn \\ & \quad
+\sin 2\theta\, \sin (4 \phi_h - \phi_R- \phi_S) \,{\cal I}\Biggl[\Biggl(
\frac{\vec{k}_T^2\lf[2(\vec{p}_T \cdot \h)^2- \vec{p}_T^{\, 2}\rg]}{4 M^2 M_h^2} - 
2 (\vec{k}_T \cdot \h) \nn \\ & \quad \times
\frac{\lf[ 4(\vec{p}_T\cdot\h)^2(\vec{k}_T\cdot\h)-2(\vec{p}_T\cdot\h)
(\vec{p}_T\cdot\vec{k}_T)-\vec{p}_T^{\, 2}(\vec{k}_T\cdot\h) \rg]}{4 M^2 M_h^2}\Biggr) 
\, h_{1T}^{\perp} \, \lf(-\frac{M_h}{|\vec{k}_T|}H_{1,LT}^{\perp} \rg)\Biggr]
\nn \\ & \quad
+\sin^2 \theta\, \sin (5 \phi_h - 2 \phi_R-\phi_S) \,{\cal I}\Biggl[\Biggl(
\frac{2 \vec{k}_T^2 (\vec{k}_T \cdot \h)
\lf[2(\vec{p}_T \cdot \h)^2- \vec{p}_T^{\, 2}\rg]}{4 M^2 M_h^3}
+ \lf[\vec{k}_T^2 - 4 (\vec{k}_T \cdot \h)^2\rg] 
\nn \\ & \quad \times
\frac{ \lf[ 4(\vec{p}_T\cdot\h)^2(\vec{k}_T\cdot\h)-2(\vec{p}_T\cdot\h)
(\vec{p}_T\cdot\vec{k}_T) -\vec{p}_T^{\, 2}(\vec{k}_T\cdot\h) \rg]}{4 M^2 M_h^3}\Biggr) \,
h_{1T}^{\perp} \, \lf(-\frac{2 M_h^2}{|\vec{k}_T|^2}H_{1,TT}^{\perp} \rg)\Biggr]
\Biggr\}, \nn
\end{align}


\subsubsection{Polarized lepton beam and transversely polarized target} 

\begin{equation} 
\begin{split}
\label{e:crosssecLT}
d^8\!\sigma_{LT}  
&=
\sum_{a} \frac{\alpha^2 e_a^2}{2 \pi s x y^2}\,\lambda_e 
\,\sst\,C(y)\,\Biggl\{ 
\cos(\phi_h-\phi_{S}) \,{\cal I}\left[\frac{\vec{p}_T\cdot\h}{M}\,
     g_{1T} \, \left(\frac{1}{4} D_{1,OO}^s + \frac{3}{4} D_{1,OO}^p\right) \right]
\\ & \quad +
+ \frac{1}{3}\lf(3 \cos^2{\theta} -1\rg)
 \cos(\phi_h - \phi_S)
\,{\cal I}\left[\frac{\vec{p}_T\cdot\h}{M}\,
     g_{1T} \, \left(\frac{3}{4} D_{1,LL} \right) \right]
\\ & \quad
-\sin 2\theta \cos(\phi_R - \phi_S)
	\,{\cal I}\left[\frac{(\vec{p}_T \cdot \vec{k}_T)}{2 M M_h} 
		\,g_{1T} \, \lf(-\frac{M_h}{2|\vec{k}_T|}D_{1,LT}\rg)\right]
 \\ & \quad
-\sin 2\theta \cos(2 \phi_h - \phi_R - \phi_S)
	\,{\cal I}\left[\frac{2(\vec{p}_T \cdot \h)(\vec{k}_T
 \cdot \h) - \vec{p}_T\cdot \vec{k}_T}{2 M M_h}
		\,g_{1T} \, \lf(-\frac{M_h}{2|\vec{k}_T|}D_{1,LT}\rg)\right]
 \\ & \quad
-\sin^2 \theta \cos(\phi_h - 2 \phi_R + \phi_S)
	\,{\cal I}\left[\frac{2 (\vec{k}_T\cdot\h) (\vec{p}_T\cdot\vec{k}_T)
	- \vec{k}_T^2(\vec{p}_T\cdot\h)}{2 M M_h^2} 
		\,g_{1T} \, \lf(-\frac{M_h^2}{|\vec{k}_T|^2}D_{1,TT} \rg)\right]
 \\ & \quad
-\sin^2 \theta \cos(3 \phi_h - 2 \phi_R - \phi_S)
	\,{\cal I}\Biggl[\Biggl(\frac{4(\vec{k}_T\cdot\h)^2(\vec{p}_T\cdot\h)
	-2(\vec{k}_T\cdot\h)(\vec{p}_T\cdot\vec{k}_T)}{2 M M_h^2}
 \\ & \quad 
	-\frac{\vec{k}_T^2(\vec{p}_T\cdot\h)}{2 M M_h^2}\Biggr)
		\,g_{1T} \, \lf(-\frac{M_h^2}{|\vec{k}_T|^2}D_{1,TT} \rg)\Biggr]
-\sin 2\theta \cos(\phi_R - \phi_S)
	\,{\cal I}\left[\frac{(\vec{p}_T \cdot \vec{k}_T)}{2 M M_h} 
		\,f_{1T}^{\perp} \, \lf(\frac{|\vec{R}|}{2 M_h} G_{1,LT}^{\perp} \rg)
		\right] \\ & \quad
+\sin 2\theta \cos(2 \phi_h - \phi_R - \phi_S)
	\,{\cal I}\left[\frac{2(\vec{p}_T \cdot \h)(\vec{k}_T
 \cdot \h) - \vec{p}_T\cdot \vec{k}_T}{2 M M_h}
		\,f_{1T}^{\perp} \, \lf(\frac{|\vec{R}|}{2 M_h} G_{1,LT}^{\perp} \rg)
		\right] \\ & \quad
-\sin^2 \theta \cos(\phi_h - 2 \phi_R + \phi_S)
	\,{\cal I}\left[\frac{2 (\vec{k}_T\cdot\h) (\vec{p}_T\cdot\vec{k}_T)
	- \vec{k}_T^2(\vec{p}_T\cdot\h)}{2 M M_h^2} 
		\,f_{1T}^{\perp} \, \lf(\frac{|\vec{R}|}{2|\vec{k}_T|} G_{1,TT}^{\perp} 
		\rg)\right] \\ & \quad
+\sin^2 \theta \cos(3 \phi_h - 2 \phi_R - \phi_S)
	\,{\cal I}\Biggl[\Biggl(\frac{4(\vec{k}_T\cdot\h)^2(\vec{p}_T\cdot\h)
	-2(\vec{k}_T\cdot\h)(\vec{p}_T\cdot\vec{k}_T)}{2 M M_h^2}
 \\ & \quad 
	-\frac{\vec{k}_T^2(\vec{p}_T\cdot\h)}{2 M M_h^2}\Biggr)
		\,f_{1T}^{\perp} \, \lf(\frac{|\vec{R}|}{2|\vec{k}_T|} G_{1,TT}^{\perp} 
		\rg)\Biggr] \Biggr\} \; .
\end{split}
\end{equation} 

The pure $p$-wave sector of the previous cross sections
 corresponds to the results of spin-1 production presented 
in Refs.~\cite{Bacchetta:2000jk,thesis}, once we apply the following
identifications
\begin{align}
\frac{3}{4} D_{1,OO}^p &= D_1, &  \frac{3}{4} D_{1,LL} &= D_{1LL}, \nn \\
-\frac{M_h}{2|\vec{k}_T|}D_{1,LT}&= D_{1LT}, & 
		-\frac{M_h^2}{|\vec{k}_T|^2}D_{1,TT}&= D_{1TT}, \nn \\
\frac{|\vec{R}|}{2 M_h} G_{1,LT}^{\perp} &= G_{1LT}, &  
		\frac{|\vec{R}|}{2|\vec{k}_T|} G_{1,TT}^{\perp} &= G_{1TT},\nn \\
\frac{3}{4} H_{1,OO}^{\perp\,p} &= H_1^{\perp}, 
		&  \frac{3}{4} H_{1,LL}^{\perp} &= H_{1LL}^{\perp}, \\
-\frac{|\vec{R}|}{2 M_h}H_{1,LT}^{\open}&= H_{1LT}, & 
		-\frac{M_h}{|\vec{k}_T|}H_{1,LT}^{\perp}&= H_{1LT}^{\perp},\nn \\
 -\frac{|\vec{R}|}{|\vec{k}_T|}H_{1,TT}^{\open}&= H_{1TT}, & 
		-\frac{2 M_h^2}{|\vec{k}_T|^2}H_{1,TT}^{\perp}&= H_{1TT}^{\perp}.
\nn
\end{align} 
Note, however, that while the functions on the left hand side contain a
dependence on $z$ as well as on the invariant mass $M_h^2$, the functions on
the right hand side depend only on $z$: it is required to assume that the
spin-1 functions behave as resonances (Breit-Wigner shapes) in the invariant 
mass.


\bibliographystyle{apsrev}
\bibliography{mybiblio}

\end{document}